\documentclass[twocolumn]{aastex631}
\usepackage{amsmath}
\usepackage{url}
\usepackage{multirow}
\usepackage{mathtools, nccmath}
\shorttitle{Resolution Limits of Astronomical Periodograms}
\shortauthors{Victor Ramirez Delgado, et al.}
\begin{document}

\title{The Rayleigh Criterion: Resolution Limits of Astronomical Periodograms}

\author[0000-0001-8183-459X]{Victor Ramirez Delgado}
\affiliation{Department of Physics and Astronomy, University of Delaware, 217 Sharp Lab, Newark, DE 19716, USA} 
\correspondingauthor{Victor Ramirez Delgado}
\email{viclrd@udel.edu}

\author[0009-0005-7084-2160]{Joan Sebastián  Caicedo Vivas}
\affiliation{Universidad del Valle, Cali, Valle del Cauca, Colombia}

\author[0000-0002-8796-4974]{Sally Dodson-Robinson} 
\affiliation{Department of Physics and Astronomy, University of Delaware, 217 Sharp Lab, Newark, DE 19716, USA} 
\affiliation{Bartol Research Institute, Sharp Lab, 104 The Green, Newark, DE, 19716, USA} 

\author[0000-0003-3996-773X]{Charlotte Haley} 
\affiliation{Mathematics and Computer Science Division, Argonne National Laboratory, Lemont, IL}



\begin{abstract}

The Rayleigh criterion determines the resolution limit of a periodogram, which is the minimum frequency separation required to barely resolve two sinusoids. Failing to consider the Rayleigh criterion may result in incorrect interpretations of long-period signals or spurious claims that two closely spaced periodogram peaks represent two distinct physical processes. Resolution considerations can help astronomers decide which periodogram peaks truly represent oscillatory signals, a question that is of great importance for exoplanet detection. We demonstrate how applying the Rayleigh criterion can help observers avoid false positive detections caused by uneven observing cadence or insufficient observing time baseline. We present three synthetic datasets that showcase the importance of the Rayleigh criterion in interpreting the generalized Lomb-Scargle and Bayesian periodograms. Our synthetic datasets illustrate (1) a single oscillation with a split Lomb-Scargle periodogram peak resulting from uneven observing cadence can be mistaken for two oscillations if the Rayleigh criterion is neglected, (2) oversampling a periodogram's frequency grid does not improve resolution, and (3) observing time baseline requirements for resolving two closely spaced oscillations. We use the Rayleigh criterion to revisit detections of planets, stellar activity, and differential rotation from four published datasets. We show that the frequency separation between planet 55~Cnc~d and the activity cycle is too small to distinguish the two phenomena based on published radial velocities (RVs) alone. Likewise, the contested 4970-day planet orbiting HD~99492 cannot be statistically separated from zero frequency. We determine that a cubic polynomial better explains the long-term RV variability of Barnard's star than a sinusoid model. Finally, our re-analysis of {\it Kepler} observations of two active stars shows that the signals previously attributed to differential rotation can be modeled by a Gaussian process with a single quasiperiodicity. This work demonstrates the importance of considering Rayleigh resolution when constructing a time-domain model.

\end{abstract}



\section{Introduction}\label{sec:intro}

The Rayleigh criterion is a widely known concept in optics
which quantifies the resolvability of two distinct light sources based on their spatial separation. According to the Rayleigh criterion,
two light sources are just resolvable when the center of the diffraction pattern of one is directly over the first minimum of the diffraction pattern of the other \citep{rayleigh1879}. Both the concept of resolution and the Rayleigh criterion apply to exploratory Fourier analysis of the type used in radial velocity (RV) planet hunting, asteroseismology, and studies of stellar rotation. In a power spectrum estimate, the analogues of the two light sources are two different oscillations with frequencies $f_1$ and $f_2$. If the frequencies are almost equal, such that $\lvert f_1 - f_2 \rvert \ll f_1$, 
the oscillations might become indistinguishable.

Similarly to how diffraction patterns occur when light passes through an 
aperture, 
discrete sampling of a continuous function over a finite time baseline 
causes spectral leakage, in which power from a process with frequency $f_0$ shows up at frequencies far from $f_0$ \citep{smith2011spectral}.
This leakage is what limits the resolution of power spectrum estimators.\footnote{Note that we differentiate between the words estimate and estimator. An estimator is a rule by which we calculate a parameter from a sample, while an estimate is the actual value calculated from that sample.} The simplest power spectrum estimator is the standard \citet{schuster1898} periodogram for time series with unit sampling ($\Delta t = t_{n+1} - t_n = 1$, where $n= 0,\ldots, N-1$ is a time index), which is defined as 
\useshortskip
\begin{equation}
\hat{S}^P(f) = \frac{1}{N} \left| \sum_{t=0}^{N-1} y(t) \exp(-2 \pi i f t) \right|^2.
\label{eq:schuster_periodogram}
\end{equation}
In Equation \eqref{eq:schuster_periodogram}, $y(t)$ is the time series, $f$ is the frequency, and $N$ is the number of observations. Formally, for two periodogram peaks to be statistically distinguishable from one another,
they must be separated by twice 
the Rayleigh resolution $\mathfrak{R}$, which is defined as
\begin{equation}
    \mathfrak{R} = \frac{1}{T},
    \label{eq:rayleigh_total_observation}
\end{equation}
where $T = t_{N-1} - t_0$ is the observing time baseline, or duration \citep[e.g.][]{godin72, Dalsgaard1982, naidu95}. 

The Rayleigh resolution is an important (though underappreciated) consideration in exoplanet searches. A non-oscillatory component of the time series has $f_1 = 0$, which means the minimum oscillation frequency that can be detected in a periodogram is given by $f_2 = 2 \mathcal{R}$. Yet there are many exoplanets discovered in Doppler searches with reported periods $P > T/2$ that have not been confirmed by either astrometry or direct imaging; a partial list of these cases is shown in Table \ref{table:SignalsRR}. Even if a mathematical model such as a Keplerian appears to fit the data, a time baseline of one (apparent) period might not be enough to guarantee that the signal under investigation is truly periodic, or if it is, that its period has been measured accurately \citep{Horne1986}. Planet hunters must also be cautious about shorter-period terrestrial planet candidates, as \cite{Vanderburg2016} showed that stellar rotation and its harmonics can have similar frequencies to the orbits of planets in M-dwarf habitable zones.

\begin{table*}[ht]\label{table:SignalsRR}
\centering
\caption{Signals reported in the literature where the reported period is longer than the time baseline and therefore does not satisfy the Rayleigh Criterion.}
\begin{tabular}{|c|c|c|c|c|c|}
\hline
\textbf{Star ID} & \textbf{Signal} & \textbf{Period (days)} & \textbf{$T$ (days)}&  \textbf{$f/\mathfrak{R}$} & \textbf{Reference} \\
\hline
{HR 5183} & HR 5183 b & $27400.0$ & 8213 & 0.304&  \cite{Blunt2019} \\
\hline
{HD 213472} & HD 213472 b & $16700$ & $6763$ & 0.405&  \cite{Rosenthal2021} \\
\hline
{47 UMa} & 47 UMa d & $14002$ &  7175 & 0.512&  \cite{gregory10} \\
\hline 
{HD 92788} & HD 92788 c & $9857$& 5877 & 0.596& \cite{Wittenmyer2019} \\
\hline 
{HD 89744} & HD 89744 c & $6974$ & 5186 & 0.744&  \cite{Wittenmyer2019} \\
\hline 
{HD 50499} & HD 50499 c & $8620$ & 2524 & 0.293& \cite{Rickman2019} \\
\hline 
\end{tabular}
\end{table*}

The ability to estimate stellar parameters using asteroseismology is also limited by the Rayleigh resolution. According to \citet{Dalsgaard1982}, \citet{claverie79} were the first to resolve the sun's $67 \mu$Hz separation of high-$n$, low-$\ell$ oscillation modes because their time series of disk-integrated line of sight velocities had sufficient time baseline, whereas previous investigations did not. \citet{claverie79} suggested that there might be a metal abundance gradient between the deep convection zone and the surface, a possibility that was verified by solar structure models \citep[e.g.][]{chaboyer95, brun02, baturin15}. Oscillation mode spacing is one of the primary constraints on solar metallicity \citep{basu08}. In addition to blends between two or more oscillation modes, \citet{Aerts2010} caution that periodogram peaks caused by true oscillations may also blend with noise peaks, increasing the frequency-estimate errors.



Additional complications result from the fact that ground-based astronomical time series are unevenly sampled due to daytime and seasonal gaps, weather-related interruptions, and telescope scheduling. The possibility of observing cadence-induced false positives motivates using the Rayleigh criterion to validate all power spectrum peaks by ensuring that their frequencies are statistically distinguishable each other and from zero. A similar complication for unevenly spaced time series is there is no formal definition of the highest frequency that the time series probes (Nyquist frequency), although some observers have proposed approximations to the Nyquist frequency \citep[e.g.][]{Koen2006}.

In their study of frequency resolution in exploratory Fourier analysis of astronomical data, \citet{Loumos1978} investigated multiple sinusoids with closely spaced frequencies similar to the pulsations reported in $\delta$ Scuti and $\beta$ Cepheid stars. They highlighted the fact that Fourier analysis can yield misleading results for short observing time baselines. 
In experiments involving two sinusoids with frequencies $f_1$ and $f_2$ sampled at uniform observing cadence, \citet{Loumos1978} found that the resulting Schuster periodograms contained two distinct peaks at the correct frequencies only when $\lvert f_1 - f_2 \rvert \ga 1.5 \mathfrak{R}$. Their results suggest that two signals become resolvable once the difference between their frequencies is greater than the frequency separation between the main lobe and the first side lobe in the Fej\'{e}r kernel $F_N$: 
\begin{equation}
F_N (f) =\frac{1}{N+1} \frac{\sin^2\left((N+1)f/2\right)}{\sin^2\left(f/2\right)}.
\label{eq:spectra_window}
\end{equation} 
A peak in a Schuster periodogram is actually a central lobe of the Fej\'{e}r kernel, which is the spectral window produced by a finite-length, otherwise untapered time series (see \citet{harris78} for more information about spectral windows). 

Another astronomical study of periodogram resolution was performed by \cite{kovacs1981}, who also used numerical experiments to investigate the frequencies at which peaks appear in the Schuster periodogram, compared with the true frequencies of the input signals. They tested input functions of the form
\begin{align}
    y_1(t) = A \sin\left(2\pi f t + \phi\right)\\
    y_2(t) = A_1\sin\left(2\pi f_1 t + \phi_1\right) + A_2\sin\left(2\pi f_2 t + \phi_2\right),
    \label{eq:kovacs_test}
\end{align}
where $\phi$ is a phase offset.
Because each sinusoid manifests in the periodogram as a Fej\'{e}r kernel,
$y_2(t)$ yields a periodogram in which the measured frequencies $f_1^{\prime}$ and $f_2^{\prime}$ deviate from the true frequencies $f_1$ and $f_2$
due to the interference of Fej\'{e}r kernels centered at $f_1$,  $\sim (f_1+f_2)/2$, and $f_2$. 
\cite{kovacs1981}'s numerical tests found that the deviations $f_1^{\prime} - f_1$ and $f_2^{\prime} - f_2$ depend on the phases and amplitude ratios, and concluded that
\begin{equation}
    |f_1 - f_2| \simeq 1.45  \mathfrak{R}
\end{equation} 
\noindent determines the smallest resolvable frequency separation.


Signal processing literature commonly cites $2 \mathfrak{R}$ as the resolution limit of a 
Schuster periodogram \citep{Thomson2007SolarMS, smith2011spectral}, albeit with some exceptions (for example, \citet{Abe2004} quote 
a resolution limit of $1.44 \mathfrak{R}$, similar to the value from \cite{kovacs1981}, in a paper on audio engineering). The study by \cite{Dalsgaard1982} on the periodogram of solar-like oscillations showed that a separation of $2\mathfrak{R}$ is accurate to resolve all relative phases and modes. The $2 \mathfrak{R}$ resolution limit also appears in numerical ecology: \citet[Chapter 12]{legendre2012numerical} explain how frequency resolution determines that the maximum period that can be safely investigated in time series is $P_{\mathrm{max}} = 1 / (2 \mathfrak{R}) = T/2$. Similarly, \cite{Black1982}'s study on detecting exoplanets through astrometry showed that if the oscillation period is greater than $P_{\mathrm{max}}$, it can lead to significant errors in disentangling orbital motion from proper motion.
Studies in oceanography and tidal analysis also put special emphasis on frequency resolution \citep[e.g.][]{thomson2014}.
Furthermore, \cite{fes2014} mention that unequally spaced time series (nonuniform $\Delta t$) tend to produce ``over-optimistic'' diagnostics of the Rayleigh criterion, meaning that the Lomb-Scargle periodogram can in practice have poorer resolution than the formal value $2\mathfrak{R}$. 


This paper defines the exploratory Fourier analysis context in which resolution must be considered, with particular emphasis on Doppler planet searches. We use numerical experiments based on synthetic data to build physical intuition about Rayleigh resolution, then present case studies of archival data to illustrate the practical use of the Rayleigh criterion. In Sect. \ref{sub:planets}, we explain how the periodogram is used---particularly in planet hunting---and describe how resolution informs time-domain RV model selection.
In Sect. \ref{sec:ray_crit} we define the Rayleigh criterion and two corollaries.
In Sect. \ref{sec:syndata} we demonstrate how the Rayleigh criterion can help observers identify Lomb-Scargle periodogram artifacts produced by uneven observing cadences and short time baselines. We also show that the same Rayleigh criterion applies to the Bayesian generalized Lomb-Scargle periodogram \citep{mortier15}.
In Sect. \ref{sec:results} 
we apply the Rayleigh criterion to RV observations of 55~Cnc, HD 99492, and Barnard's Star, as well as {\it Kepler} light curves of KIC~891916 and KIC~1869783. Finally, Sect. \ref{sec:conclusions} summarizes our results and presents recommendations for applying the Rayleigh criterion to Lomb-Scargle periodograms.

\section{The Role of the Periodogram} \label{sub:planets}


Doppler planet searches, asteroseismology, and studies of star rotation (among others) rely heavily on the 
Lomb-Scargle periodogram \citep{lomb1976, scargle1982} and its extensions, which include the generalized Lomb-Scargle periodogram \citep{Zechmeister2009}, Kepler periodogram \citep{otoole09, gregory16}, multiharmonic periodogram \citep{baluev09}, residual periodogram \citep{angladaescude12}, and Bayesian / stacked Bayesian generalized Lomb-Scargle periodograms \citep{mortier15, mortier2017}.\footnote{Other period-search tools include compressed sensing \citep{hara2017}, which adds the assumption of sparsity in the frequency domain, and Welch's method, in which periodograms computed from tapered time series segments are averaged to produce a power spectrum estimator with reduced variance \citep{welch67, dodson2022}.}
Once periodic signals have been identified, the planet hunter can construct an RV model with free parameters that describe the planets plus any stellar signals with high enough amplitudes not to be subsumed by the instrumental noise.
A typical model consists of one or more Keplerian orbits,
a Gaussian processes (GP) to describe quasiperiodic rotation and/or magnetic activity cycles, and sometimes a red noise parameterization \citep[e.g.][]{angladaescude13, tuomi13, rajpaul15, yu17, faria22, suarezmascareno2023}.

The periodogram's role as an exploratory tool is powerfully demonstrated at the model construction phase. Incorrect physical inferences come from leaving out necessary model components, such as the inaccurate mass measurements of CoRoT-7 b and c that resulted from neglecting star rotation and activity \citep{haywood14}. Including spurious model components, such as the planet with a 233-day period orbiting Barnard's star that was later challenged by \citet{gonzaleshernandez2024}, also leads to errors in physical understanding.



Our motivation in writing this paper is to reduce mistakes in exploratory Fourier analysis, which guides the observer in constructing a model for the time series. One preventable class of mistakes comes from neglecting to consider periodogram resolution when searching for periodic signals. Before we begin our mathematical exploration of Rayleigh resolution, we will explain the periodogram applications in which resolution is relevant (Sect.\ \ref{sec:periodogram}) and discuss the resolution-related pitfalls presented by iterative fitting and subtraction of Keplerians or sinusoids (Sect.\ \ref{subsec:iterative_prewhitening}).

\subsection{Power Spectrum Estimation vs.\ Oscillation Frequency Estimation}
\label{sec:periodogram}

Here we discuss two distinct uses of the periodogram: (1) estimating the power spectrum of a stationary time series (spectral analysis) and (2) estimating oscillation frequencies (harmonic analysis). The goal of (1) is to partition the time series variance into its oscillatory components so that the amplitude at each frequency gridpoint has physical meaning \citep[Chapter 4]{shumwaystoffer}. Used in a spectral analysis context, Schuster and Lomb-Scargle periodograms are nonparametric, meaning they presuppose no particular model for the data. Instead, they are exploratory tools that can reveal the types of signals that are present in the data. For example, a power spectrum estimate from a {\it Kepler} time series may show p-mode oscillations, granulation, rotation, evolution of active regions, and one or more transiting planets. In RV planet hunting, observers attempt to zero out the p-mode contributions with carefully chosen exposure times, but must contend with magnetic activity cycles when time baselines are long. A nonparametric power spectrum estimate is usually not an end unto itself, but rather a guide for selecting an appropriate time-domain model with free paramaters that are connected to the underlying physics \citep[though sometimes models are fitted in the frequency domain, as in asteroseismology;][]{Aerts2010}.

On the other hand, harmonic analysis is a frequency domain model-fitting process conducted under the null hypothesis that the time series records an oscillation. The accompanying assumption about the power spectrum is that it includes a delta function at frequency $f_{\delta}$, where $f_{\delta}$ is the free parameter of the model. The frequency of the highest periodogram peak then serves as an estimator for $f_{\delta}$. If the observer specifies a particular time series covariance structure \citep[e.g.\ correlated noise,][]{baluev13, delisle20}, the frequency-domain model contains more free parameters. The Bayesian generalized Lomb-Scargle periodogram \citep{mortier15} is a harmonic analysis tool that recasts the generalized Lomb-Scargle periodogram \citep{Zechmeister2009} as a probability distribution of $f_{\delta}$.

The Rayleigh resolution limit described in Sect. \ref{sec:intro} does not apply to harmonic analysis, as there are ways to extract estimates of $f_{\delta}$ with better precision than $\mathfrak{R}$ \citep[e.g.][]{capon69, kaplun23}. For example, in a time series with uncorrelated errors and uniform time sampling (constant $\Delta t$), \citet{montgomery1999derivation} find that the error on the estimated oscillation frequency follows $\sigma(f_{\delta}) \propto \mathfrak{R}^{3/2}$ \citep[see also][]{walker1971estimation}. \cite{Schwarzenberg-Czerny1991} determine that the same uncertainty in the frequency estimate is applicable to unevenly spaced observations if the average timestep is much less than the noise correlation time and the time sampling is not taken at similar intervals to the signal's period (i.e.\ $\overline{\Delta t} \not\approx P$). But the Rayleigh limit is extremely important for spectral analysis, which is a step that cannot be avoided without introducing the potential for fitting the wrong model.

Returning to the RV planet hunting example, the Lomb-Scargle periodogram will not have a peak associated with rotation if the star's rotation axis points toward the sun. In such cases, it would not make sense to include a Gaussian process with a quasiperiodic kernel in the time-domain model \citep[e.g.][]{haywood14, angus18}. Magnetic activity will not manifest in the RVs if the star is in a Maunder minimum \citep{wright04}---or the star could have three distinct activity cycles, as in $\epsilon$~Eri \citep{fuhrmeister23}. By definition, an RV planet hunter does not know {\it a priori} how many Keplerian signals the time series will record, which makes harmonic analysis suboptimal for planet detection. A nonparametric power spectrum estimate is an indispensable tool for selecting a time-domain model framework that truly represents the underlying physics. 

While parametric and semi-parametric models are valuable when their underlying assumptions are satisfied\footnote{We often think of the quote ``Essentially, all models are wrong, but some are useful'' from \citet{box86}.}, they should be preceded by Rayleigh resolution-informed, nonparametric Fourier analysis so as to avoid introducing spurious model components or leaving out necessary ones. 

\subsection{Finding Multiple Oscillations with Iterative Fitting and Subtraction}
\label{subsec:iterative_prewhitening}

One difficulty in detecting weak oscillations using a periodogram is that spectral leakage from strong (usually low-frequency) components can produce spurious power near the frequencies of interest \citep[in fact,][go so far as to suggest that spectral leakage renders the periodogram ineffective for detecting periodic signals in unevenly spaced astronomical time series]{Vio2013}.
The weak signals are only uncovered after fitting and subtracting a model of the strong signal. Iterative fitting and subtraction of Keplerians and sinusoids\footnote{While iterative fitting and subtraction of periodic models is often called prewhitening in the astronomical literature, statisticians use the term prewhitening to describe removal of a model for the power spectrum continuum.} are therefore mainstays of frequency-domain methodology. The observer finds the frequency of the highest peak in a Lomb-Scargle periodogram, subtracts a periodic model with that frequency from the time series, computes a periododogram of the residuals, and repeats either a specified number of times \citep[e.g.][]{kepler_stars, reinhold15} or until some quantitative stopping criterion is reached \citep[e.g.][]{blomme11, hatzes18, dornwallenstein19}.

As discussed by \citet{jenkins14}, iterative sinusoid or Keplerian subtraction can be dangerous because an error at any step will propagate into subsequent steps.
Applying the Rayleigh criterion can help eliminate problems with the iterative procedure. In Sect.\ \ref{subsec:diffrot} we show an example in which potentially spurious sinusoids were introduced into Fourier-series models of star rotation because the Rayleigh criterion was not applied at each fitting and subtraction step. Now is the time to improve the iterative methodology, before the launch of exoplanet time-domain missions such as Twinkle \citep{stotesbury22}, Atmospheric Remote-sensing Infrared Exoplanet Large-survey \citep[ARIEL;][]{ariel}, and PLAnetary Transits and Oscillations of stars \citep[PLATO;][]{plato2024}.

\section{Rayleigh Criterion} \label{sec:ray_crit}

First we consider the frequency resolution required to distinguish between the sinusoids $y_1(t) = A_1 \sin (2 \pi f_1 t + \phi_1)$ and $y_2(t) = A_2 \sin (2 \pi f_2 t + \phi_2)$. We state the \textbf{Rayleigh criterion} as 
\begin{equation}
    |f_1 - f_2| \geq C \mathfrak{R},
    \label{eq:rayleigh_crit}
\end{equation}
where $C$ is a constant of order unity. According to \citet{thomson2014}, the signals are well resolved if $1.5 < C < 2$ (see their Figure 1.2.2). \citet{BRAUN2001} suggests a more conservative Rayleigh criterion of $C = 1.5$--3, 
while \cite{kovacs1981} and \cite{Abe2004} use a more optimistic value of $C = 1.44$--1.45. 
Since the fundamental property of periodic signals is that they repeat, we argue that $C = 2$ gives the appropriate Rayleigh criterion, as it ensures that the time series contains a repeat of every part of every wave---including the beating between $y_1$ and $y_2$. (We will present experimental evidence that observers should adopt $C = 2$ in Sect. \ref{sec:syndata}.) 
We introduce two corollaries to Equation \eqref{eq:rayleigh_crit}:

\begin{itemize}

    \item \emph{Corollary 1}: For an oscillation to be detected using a periodogram, it must be distinguishable from zero frequency according to the Rayleigh criterion. Thus, the lowest frequency observable is $f_\mathrm{min} = C\mathfrak{R}$.
    
    \item \emph{Corollary 2}: Zero padding or oversampling the frequency grid does not improve the Rayleigh resolution. 
    
\end{itemize}

In the second corollary, zero padding refers to appending a sequence of zeros to the end of a time series. This technique exploits a numerical property of discrete Fourier transforms to produce a smoother periodogram, but does not cause the original time baseline to increase. Both corollaries are fundamental to this study and will be used throughout the following two sections to interpret our results. 

We use the generalized Lomb-Scargle periodogram (GLSP) \citep{Zechmeister2009} to estimate the power spectra of unevenly spaced time series. Refer to Appendix \ref{sec:gsl} for an overview of the GLSP. 

\begin{figure}

\centerline{\includegraphics[width=1\linewidth]{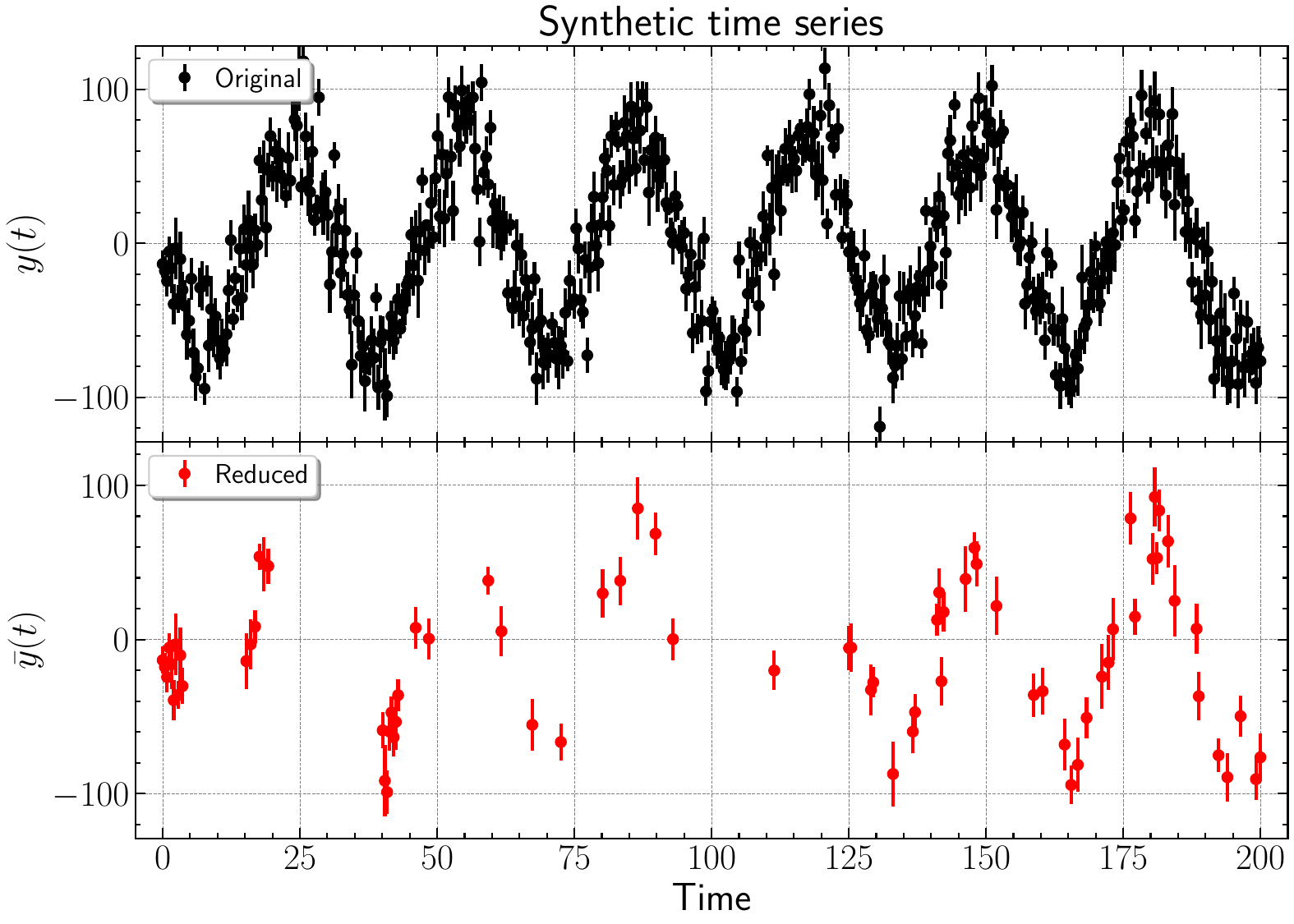}}
\caption{{\bf Top panel:} 500 samples of a single sinusoid with additive white noise are generated according to Equation \ref{eq:sinousoid} using equal sampling. {\bf Bottom panel:} 80 samples, including the first and last sample, have been selected from the time series above to create a new series with uneven observing cadence but the same nominal Rayleigh resolution.}
\label{fig:fake_data}
\end{figure}

\begin{figure*}

\centerline{\includegraphics[width=0.7\linewidth]{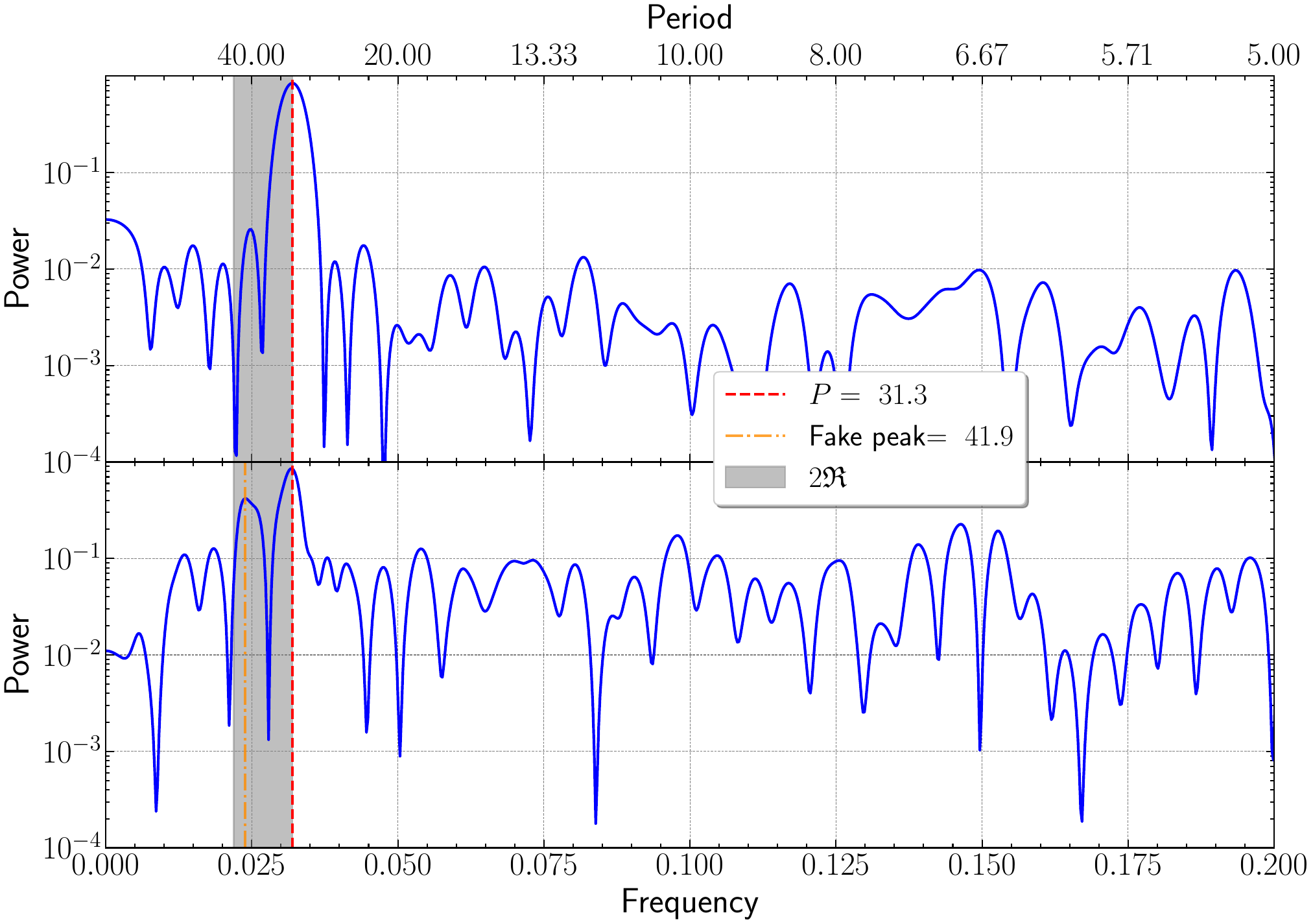}}
\caption{{\bf Top panel:} periodogram of the time series shown in the top panel of Figure \ref{fig:fake_data}, generated using Equation \ref{eq:sinousoid}. The oscillation frequency, 0.032 time units$^{-1}$ ($P = 31.3$~time units), is shown as the dashed vertical line and the shaded region illustrating the Rayleigh criterion extends to $2\mathfrak{R} = 0.01$ day$^{-1}$ from the oscillation frequency. The top x-axis shows the values in period space. {\bf Bottom panel:} GLSP of the thinned time series shown in the bottom panel of Figure \ref{fig:fake_data}. Two peaks appear instead of one, but the maximum does not occur at the true oscillation frequency, nor are the two peaks sufficiently separated to satisfy the Rayeligh criterion.}
\label{fig:periodograms_syn_data}
\end{figure*}

\section{Synthetic Datasets}\label{sec:syndata}

In this section we will apply Equation \eqref{eq:rayleigh_crit} and its two corollaries to generalized Lomb-Scargle periodograms of three synthetic datasets with known oscillatory signals. We also apply the Bayesian generalized Lomb-Scargle periodogram \citep[BGLS;][]{mortier15} to compare its sensitivity to resolution with that of the GLSP. Our numerical experiments demonstrate how considering the Rayleigh criterion can help observers avoid Fourier analysis mistakes.

\subsection{Splitting of a single periodogram peak}\label{sub:one_sig}

Here we give a toy example in which a single sinusoid plus noise produces two peaks in a GLSP. Before application of the Rayleigh criterion, one might falsely conclude that there are two sinusoids present, but on closer inspection the peaks are statistically indistinguishable.

Consider the signal
\begin{equation}
    y_n = A \sin\left(\frac{2 \pi t}{P} + \phi\right) +\mathcal{N}(0,21.22),
    \label{eq:sinousoid}
\end{equation}
\noindent where $A= 70.72$, $P = 31.33$ time units, and $\phi = 179.46^{\circ}$. $\mathcal{N}$ is a white noise term added to our function with zero mean and standard deviation $\sigma = 21.22$. We also include uncertainties 
drawn from the Gaussian distribution $\mathcal{N}(0.2 A, 4)$. 
The function $y_n$ is designed to mimic an RV time series. We begin with a realization of Equation \ref{eq:sinousoid} with 500 synthetic observations and a uniform observing cadence $\Delta t_n = t_n - t_{n-1} = 0.4$ time units, for $n = 1, 2, \ldots, N$. We then omit 425 points from the sample, while preserving the first and last data point, to obtain $y_{t_n}$ with the same Rayleigh resolution $\mathfrak{R} = 0.005$~time~units$^{-1}$ as the original. Both time series are shown in the top and bottom panels of Figure \ref{fig:fake_data}.

Figure \ref{fig:periodograms_syn_data} shows the GLSP of the full data set (top panel) and the thinned data set (bottom panel), computed using the \texttt{LombScargle} class in \texttt{astropy} version 5.2.1 \citep{astropy:2018}. Both periodograms have oversampled frequency grids with $\Delta f = f_k - f_{k-1} = \mathfrak{R}/20$. The true period is highlighted with a dashed vertical red line in both panels 
The periodograms show that for the full data set the peak centered at the true frequency $f_t = 1/31.33$~time~units$^{-1}$ has the highest amplitude, as expected. However, in the thinned series' GLSP the peak at $f_t$ is not the only significant peak; 
there is a secondary peak at $f_s = 0.0239$~time~units$^{-1}$ (orange vertical line).
The dark gray band highlights the region between $f_t$ and $f_t - 2 \mathfrak{R}$.
Since $f_s$ falls within the gray band, it is statistically indistinguishable from $f_t$; the Rayleigh criterion therefore confirms our {\it a priori} knowledge that the secondary peak is spurious.

As a limiting case, we conducted the same experiment without the white noise, i.e., we assumed the error bars were negligible. We then computed periodograms of the full dataset and the thinned dataset with the same timesteps as in the bottom panel of Figure \ref{fig:fake_data}. Comparing the results to our original experiment, we found similar behaviour to Figure \ref{fig:periodograms_syn_data}, where the peak at $f_t$ has the highest power but there is a neighboring peak at approximately the same frequency $f_s$. This experiment indicates that the spurious peak in Figure \ref{fig:periodograms_syn_data} originates from the window function of the observations and not the error bars.

We also computed the BGLS periodogram of the synthetic data in Figure \ref{fig:fake_data}, where we found that only the correct period was detected. The BGLS has $\mathcal{P}_\mathrm{BGLS} \propto e^{P_\mathrm{LS}(f)}$, where $\mathcal{P}_\mathrm{BGLS}$ is the posterior probability density as a function of frequency that a sinsuosid is detected, $P_\mathrm{LS}$ is the estimated power from the Lomb-Scargle periodogram, and the exponentiation is known to suppress alias peaks and sidelobes \citep{vanderplas2018}. In the case where there is only one true signal in the data, the BGLS is expected to find the correct frequency more easily than the GLSP. However, as we we will see below, resolution constraints must be considered in the BGLS when more than one true signal is present. In such cases the GLSP gives a better representation of the frequency domain.

 \begin{figure*}[!ht]

     \centerline{\includegraphics[width=0.7\linewidth]{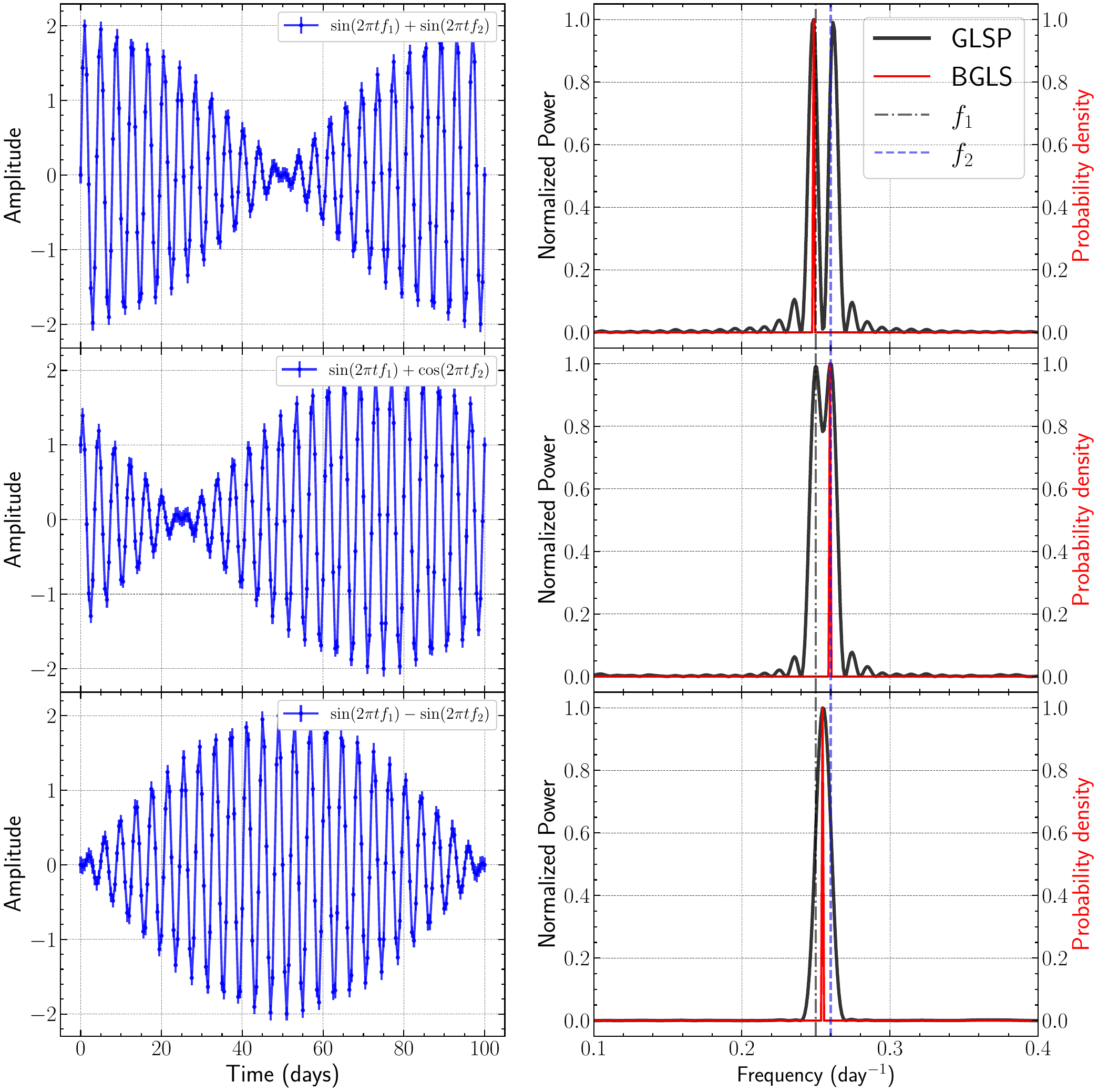}}
     \caption{\textbf{Top panel}: Time series computed according to Equation \ref{eq:zeropad} with $\phi = 0$ (left).  
     The corresponding GLSP (black) and BGLS periodogram (red) has two resolved peaks at $f_1$ and $f_2$ (right). \textbf{Center panel}: As above, but with $\phi = \pi / 2$ (left). 
     The periodogram peaks associated with $f_1$ and $f_2$ are barely resolved (right). \textbf{Bottom panel}: As above, but with $\phi = \pi$ (left).     
     The periodograms have a single peak at $(f_1 + f_2)/2$ instead of resolved peaks at $f_1$ and $f_2$ (right)}.
     \label{fig:zero_padding_peaks}
 \end{figure*}

False splitting of a single peak into a doublet can happen when there is large variation in $\Delta t_n$, as is the case in this example. We caution against naive interpretation of the GLSP without application of the Rayleigh criterion as spurious peaks can be created in the manner demonstrated. As stated in Sect.\ \ref{sub:planets}, an incorrect interpretation of a periodogram peak could lead the observer to construct a time-domain model that does not truly represent the underlying physics.

\subsection{Oversampling} \label{sub:oversample}

A GLSP or Schuster periodogram can be oversampled---i.e.\ computed on a denser frequency grid than the natural spacing of $f_k - f_{k-1} = \mathfrak{R}$---ensuring that no periodic signals are missed due to grid sparseness. In accordance with Corollary 2, we demonstrate that oversampling does not improve frequency resolution.
In this example, we consider the signal
 \begin{equation} \label{eq:zeropad}
 y(t) = \sin (2 \pi f_1 t) + \cos (2 \pi f_2 t + \phi),
 \end{equation}
in which $T = 100$ days, so that $\mathfrak{R} = 0.01$~day$^{-1}$. 
The synthetic time series has $N = 201$ data points. The frequencies were set to $f_1 = 0.25$~day$^{-1}$ and $f_2 = f_1 + \mathfrak{R}$, 
which are indistinguishable according to the Rayleigh criterion. Additionally, we assign small error bars to each data point selected from a normal distribution of the form $\mathcal{N}(0.1,0.01)$.
We compute the GLSP and BGLS of the time series for three different values of $\phi$. Each periodogram has the same frequency grid, which is oversampled such that $\Delta f = f_k - f_{k-1} = \mathfrak{R}/10$.
 
 \begin{figure*}
     \centerline{\includegraphics[width=0.85\linewidth]{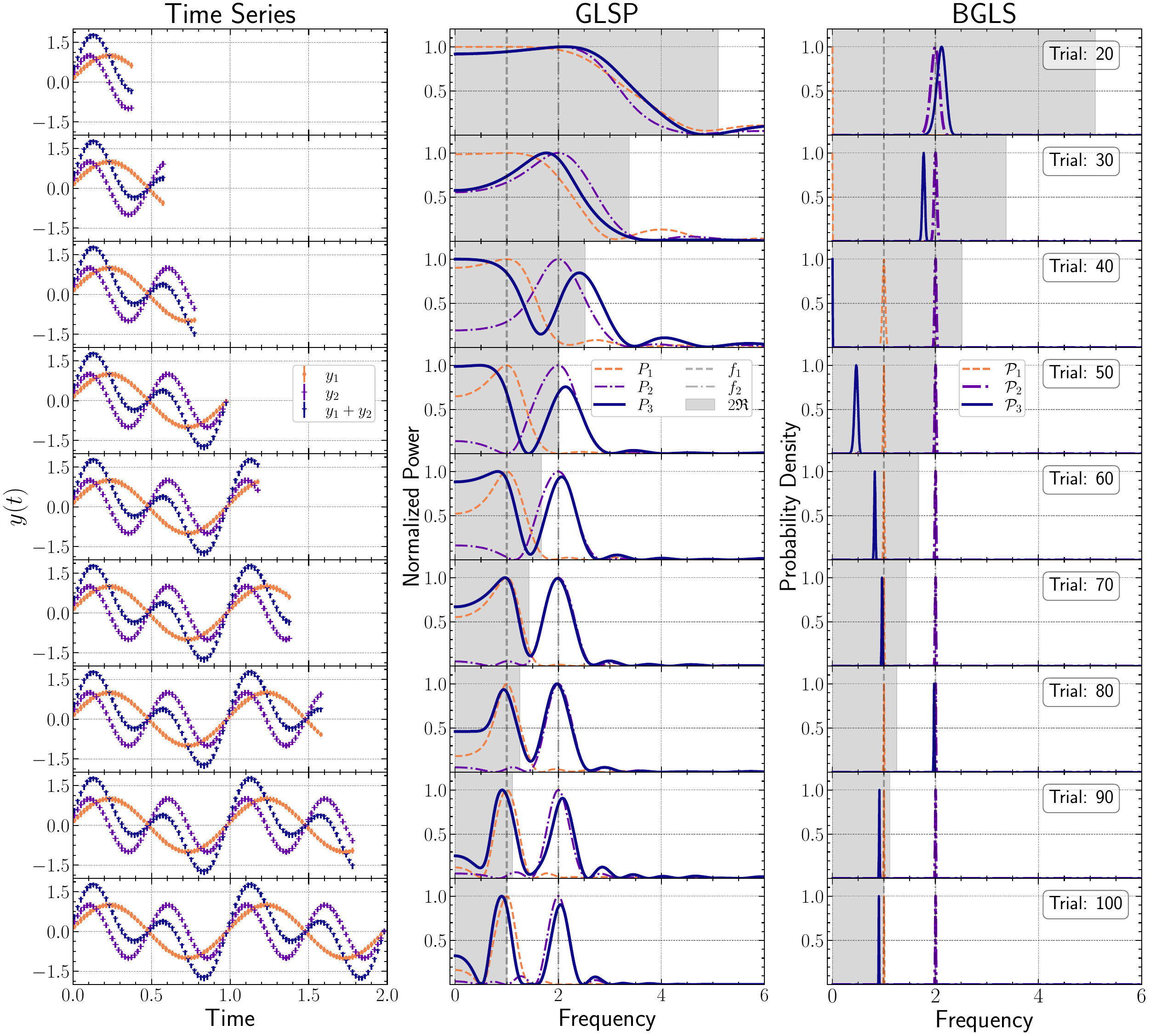}}
     \caption{{\bf Left column:} Time series for our three synthetic signals $y_1$ (dark blue dot), $y_2$ (blue cross), and $y_3$ (down trident). Each panel represents the time series at different values of $T$, the top one being evaluated at the smallest value and increasing as going down in the column. {\bf Middle column:} GLSP of the time series  on the left column, illustrating the changes of the periodogram as the signal coverage increases. The colors of each line match their time domain counterparts, and the gray band shows the $2 \mathfrak{R}$ region from the zeroth frequency. As the value of $T$ increases the shaded region recedes showing how the resolution narrows. {\bf Right column:} BGLS periodgram of the time series on the left column, following the same rules as the middle column.}
    \label{fig:snapshot}
\end{figure*}

The left panel of Figure \ref{fig:zero_padding_peaks} shows realizations of $y(t)$ for $\phi = [0, \pi/2, \pi]$, while the right panel shows their GLSPs (black) and BGLS periodograms (red). The results reveal that $\phi$ has a significant effect in resolving signals in both the GLSP and the BGLS. Consider first the GLSP: if $\phi = 0$ (Figure \ref{fig:zero_padding_peaks}, top), the observer is lucky---there are two distinct peaks at $f_1$ and $f_2$. When $\phi = \pi/2$ (Figure \ref{fig:zero_padding_peaks}, middle) the two peaks start to blend, but are still barely resolvable. However, for $\phi = \pi$ (Figure \ref{fig:zero_padding_peaks}, bottom), there is only one periodogram peak centered at the intermediate frequency $(f_1 + f_2) / 2$. The BGLS periodogram for $\phi = \pi$ has the exact same behavior: the most significant peak coincides with the GLSP's peak at $(f_1 + f_2) / 2$. It is important to note that the BGLS periodogram is meant to determine the single frequency of highest probability in the time series \citep{mortier15}, which explains why in the top panel ($\phi = 0$) there is only one peak at $f_1$. Regardless of which type of periodogram the observer uses, increasing the oversampling factor does not guarantee that the two sinusoids are resolved for all values of $\phi$.

The phase-sensitive behavior of both periodograms can be understood by considering a related situation in which a sinusoid is sampled with a cadence exactly equal to the Nyquist rate ($\Delta t = P/2$). 
If the samples fall exactly at the peaks and troughs of the wave, the observer recovers the full amplitude of the signal. If the wave is phase-shifted by 90 degrees so the samples fall at the zero crossings, the observations completely miss the oscillation and the recovered amplitude is zero. Oversampling is analogous to interpolating in the frequency domain: it smooths the power spectrum estimate, but it does not add new information.

\subsection{Varying time baseline} \label{sub:var_len}

Our last experiment with synthetic data involves testing the performance of the GLSP and BGLS periodograms for different values of the total observation time baseline $T$. 
We create timestamps for 81 time series with number of observations $N = 20, 21, \ldots, 100$; $t_0 = 0$; and equal spacing $\Delta t = 0.02$ time units. On each set of timestamps, we place the following three signals:
\begin{align}
    y_1(t) = \sin\left(2\pi t f_1\right) \label{eq:y1}\\
    y_2(t) = \sin\left(2\pi t f_2\right) \label{eq:y2} \\
    y_3(t) = y_1(t) + y_2(t),\label{eq:y3}
\end{align}
where $f_1 = 1$~time units$^{-1}$ and $f_2 = 2$~time units$^{-1}$. Once again we assign error bars to each data point following the normal distribution used in section \ref{sub:oversample}.

Figure \ref{fig:snapshot} shows the results of the time-baseline experiment. The left column illustrates
$y_1(t)$, $y_2(t)$, and $y_3(t)$ for $N = 20, 30, 40, \ldots, 100$; the middle column depicts their respective GLSPs $P_1, P_2, P_3$; and the right column shows their BGLS periodograms $\mathcal{P}_1, \mathcal{P}_2, \mathcal{P}_3$. An animated version of this figure is included in the online version of the manuscript.

Focusing on the behavior of the GLSPs shown in the middle panel of Figure \ref{fig:snapshot}, for $N < 30, T < 0.6$ time units (Trial 20), we see that all power spectrum estimates are inaccurate, with most of the power apparently concentrated at the lowest frequencies. 
Once $N = 30$ and $T = 0.6$, the signal with frequency $f_2$ is manifesting in periodograms $P_2$ and $P_3$, which have peaks near $f = f_2$. The signal with frequency $f_1$ is still indistinguishable. By $N = 40$ and $T = 0.8$, $P_2$ has a clear peak at $f = 2$ even though the time series does not yet cover two full cycles of $f_2$, but $P_3$ has a peak centered at $f = 2.4$, even less accurate than at $T = 30$.   
At $N = 50$ and $T = 1$, when the time series covers two full cycles of $f_2$ and the signal formally meets the Rayleigh criterion for separation from zero frequency, its peak in $P_3$ is finally centered at $f_2$.
As we increase the observing time baseline, $P_1$ and $P_3$ develop
peaks that start to converge toward $f_1$. Once $N = 80$ and $T = 1.6$, $P_1$ and $P_3$ have all peaks at the correct frequencies. $T = 80$ corresponds to $C = 1.6$, which is higher than the values recommended by \citet{Loumos1978} and \citet{kovacs1981}.

The BGLS periodograms shown in the right panel of Figure \ref{fig:snapshot} tell a similar story. Examining the panels corresponding to small time baselines, we see that the peaks in $\mathcal{P}_3$ land at incorrect frequencies for $N< 60, T < 1.2$. For example, when $N=40$ and $T=0.6$, the peak is lies at approximately the zeroth frequency. Once $N=70$ and $T = 1.4$ the BGLS periodogram peak of $y_3$ finds $f_2$---but after $N=90$, the single peak switches to a frequency close to $f_1$. These results show that the BGLS on its own, without a comparison to
the GLSP, can be seriously misleading
When using a Bayesian approach to computing periodograms, the observers not only have to consider frequency-domain resolution; they must also recall that the time series may contain more than one sinusoid, a situation for which the BGLS is not ideal.

The experiment illustrated in Figure \ref{fig:snapshot} shows that accurate frequency estimates are only guaranteed for oscillations with $f > 2\mathfrak{R}$, such that the Rayleigh criterion for distinguishing them from zero frequency is satisfied.
Furthermore, our synthetic datasets have only one or two sinusoids and minimal added noise. Real astronomical datasets 
may contain more than two periodic components along with multiple noise sources, diminishing the accuracy of the GLSP.
In Appendix \ref{app:timebaseline} we present additional variations of this experiment with uneven time sampling and increased noise amplitude.

\section{Archival Datasets} \label{sec:results}

In this section we apply the Rayleigh criterion to generalized Lomb-Scargle periodograms of published astronomical data. 
We investigate the long-period, low-frequency signals in radial velocities of 55~Cnc, HD~99492 and Barnard's star (Table \ref{table:orbitalparameters} includes a list of each star's periodic RV signals from the literature).
Using 55~Cnc and HD~99492 as examples, we demonstrate how the Rayleigh criterion helps observers make well-informed decisions about appropriate time-domain models. We also show that the time baseline of the \citet{ribas2018} Barnard's star radial velocities is not long enough to estimate the activity cycle frequency, and in fact the long-term variability is better fit by a cubic than a sinusoid. Finally, we reassess the detections of differential rotation in {\it Kepler} photometry of KIC 891916 and KIC 1869783.

\subsection{55~Cnc}
\label{subsec:55Cnc}

\begin{figure}

\centerline{\includegraphics[width=1\linewidth]{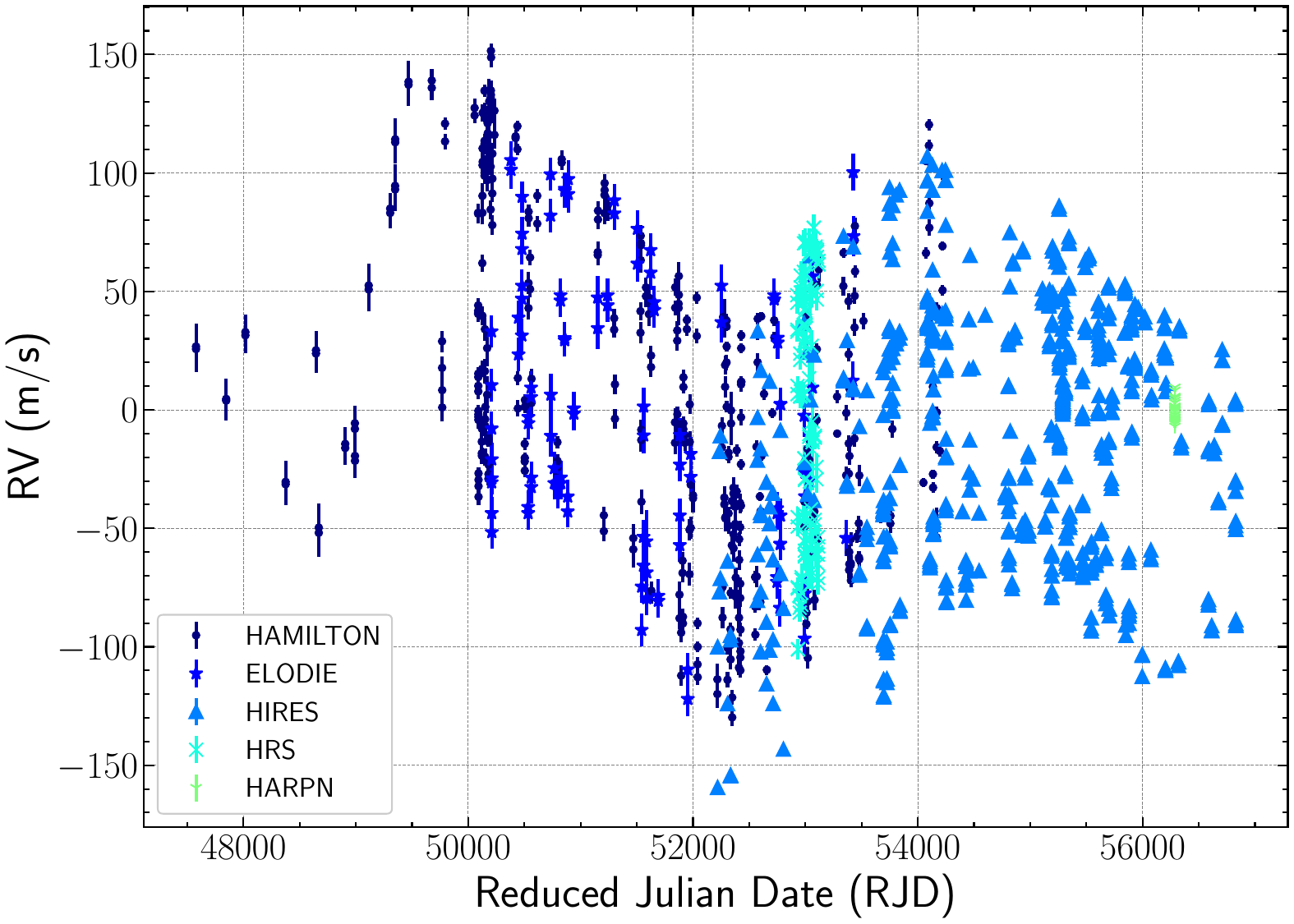}}
\caption{Radial velocities of 55~Cnc used in our analysis. The 
time series contains 1350 data points covering approximately 25 years.}
\label{fig:55cnc_rv}
\end{figure}

55~Cnc, a G8 dwarf located 12.6~pc from the sun \citep{gaia20}, has five confirmed exoplanets \citep{butler1997, marcy2002, macarthut2004, fischer2008, dawson2010}, of which planet d is a long-period gas giant. Numerous orbital periods for 55~Cnc~d have been reported in the literature, including $5360 \pm 400$ days \citep{marcy2002}, 4867 days \citep{nelson2014}, and $5574.2_{-88.6}^{+93.8}$ days \citep{bourrier2018}. 
The existence of a magnetic activity cycle is confirmed by HIRES H$\alpha$ and S-index measurements by \citet{butler2017}, periodograms of which are shown in Figure~3 of \cite{bourrier2018}. Based on a red noise model, \citet{baluev2015} estimated an activity cycle period of $12.6\pm^{2.5}_{1.0}$ yr ($4602 \pm^{913}_{365}$ days). \cite{bourrier2018} found different best-fit activity periods from different observables: 14.4 yr from photometry, $10.5 \pm0.3$ yr from the S-index, and 11.8 yr from the H$\alpha$ index. Here we assess whether published RV measurements yield a clear distinction between the periods of planet d and the magnetic activity cycle. Figure \ref{fig:55cnc_rv} shows the RV data used in our analysis, which includes observations from the Hamilton spectrograph \citep{marcy2002, fischer2008}, ELODIE \citep{naef2004}, HIRES \citep{fischer2008, butler2017}, the HRS and Tull spectrographs \citep{macarthut2004, endl12}, and HARPS/HARPS-N \citep{lopezmorales14}.\footnote{Radial velocities were provided by the Open Data module of \texttt{DACE}, \texttt{https://dace.unige.ch/dashboard/}. Data are available in the archive attached to this publication \citep{RRarchive}.} 

\begin{figure}

\centerline{\includegraphics[width =1 \linewidth]{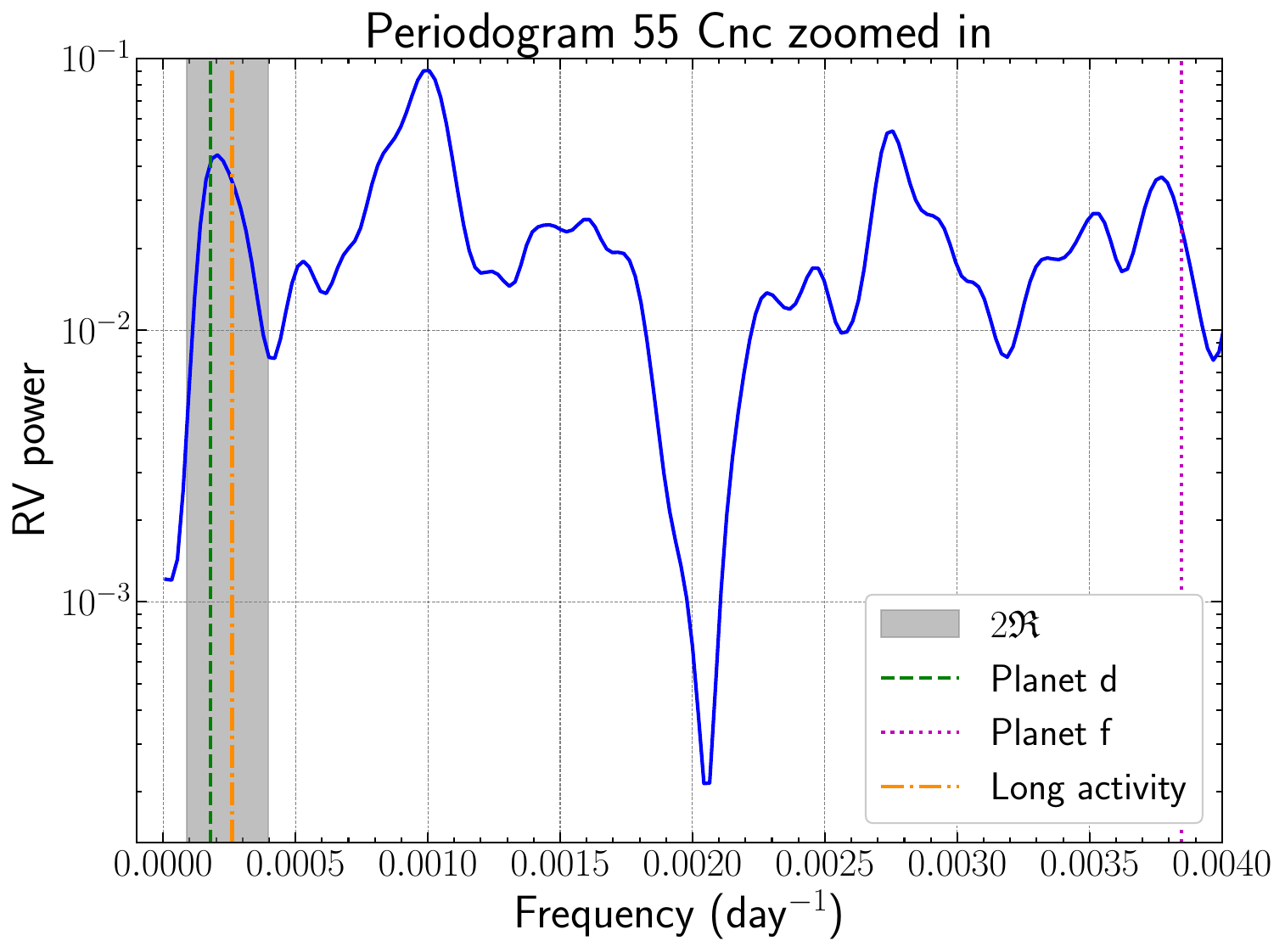}}
\caption{GLSP of 55~Cnc RVs focused in the low-frequency range.
The vertical lines show frequencies of interest: the long-term activity cycle (orange dash-dotted), planet d (green dashed), and planet f (pink dotted). The horizontal lines indicate the 10\%, 5\%, and 1\% false alarm thresholds in ascending order. 
The frequency separation between the long term activity cycle and planet d 
is less than $2\mathfrak{R}$, and in fact both frequencies fall within the same periodogram peak attributed to a period of 13 years.}
\label{fig:55cnc_per}
\end{figure}

The GLSP of the RVs in Figure \ref{fig:55cnc_per} is zoomed in on the lowest frequencies, which include planet d and the activity cycle. The gray shaded region has width $2 \mathfrak{R}$. The difference between the frequencies of planet d and the activity cycle, as reported by \cite{bourrier2018}, is $|f_d - f_{\mathrm{mag}}| = 0.75 \mathfrak{R}$. Planet d cannot be distinguished from either the activity cycle or the zero frequency based on the RV periodogram alone. We repeated our Rayleigh resolution analysis with the HIRES activity indices and found the most significant periodogram peak 
at $P \approx10.5$ years for both S-index and H$\alpha$. Since the HIRES observations have shorter time baseline than the full RV dataset, the Rayleigh criterion for estimating the activity cycle period is also not satisfied in the activity-indicator time series. However, the case for planet d is bolstered by observations from the Hubble Telescope Fine Guidance Sensors, which suggest a proper motion consistent with a 53$^\circ \pm 6.8$ inclination for planet d \citep{macarthut2004}. 

Even though there is evidence that both planet d and the activity cycle are real signals, radial velocities alone are not sufficient to separate them; other datasets must be included in the analysis. The importance of the Rayleigh criterion is that it guides the construction of the time-domain RV model, helping the observer
choose whether to include planet d plus a periodic or quasiperiodic activity cycle, the activity cycle only, or a polynomial fit that has no physical meaning but removes the long-period RV variation. Since planet d is not resolvable from the activity cycle in the frequency domain, modeling it with a Keplerian is unlikely to improve physical understanding of the 55~Cnc system. 
Either a quasiperiodic activity model or a polynomial fit would be an appropriate choice, but neither should be used to infer activity cycle parameters. 

\subsection{HD 99492}\label{sub:hd99492}

HD 99492 is a late-G/early-K type star located at 17.99 $\pm$ 1.14 pc in a binary orbit with HD 99491. \cite{marcy2005} reported the discovery of~HD 99492~b, which has an orbital period of 17.1~days and a minimum mass $M \sin i = 0.1 M_J$. \cite{meschiari2011} then reported planet candidate c, with $P = 4969.73$ days. However, \cite{kane2016} presented evidence that the 4969-day signal was actually caused by the long-term activity cycle and argued that planet c was a false positive. The most recent update on the system comes from \cite{stalport2023} with the detection of a new HD~99492~c at a period of 95.23 days. In our analysis, we revisit the data used in \cite{meschiari2011} and apply the Rayleigh criterion to the discovery of the original planet c candidate.

\begin{figure}

\centerline{\includegraphics[width=1\linewidth]{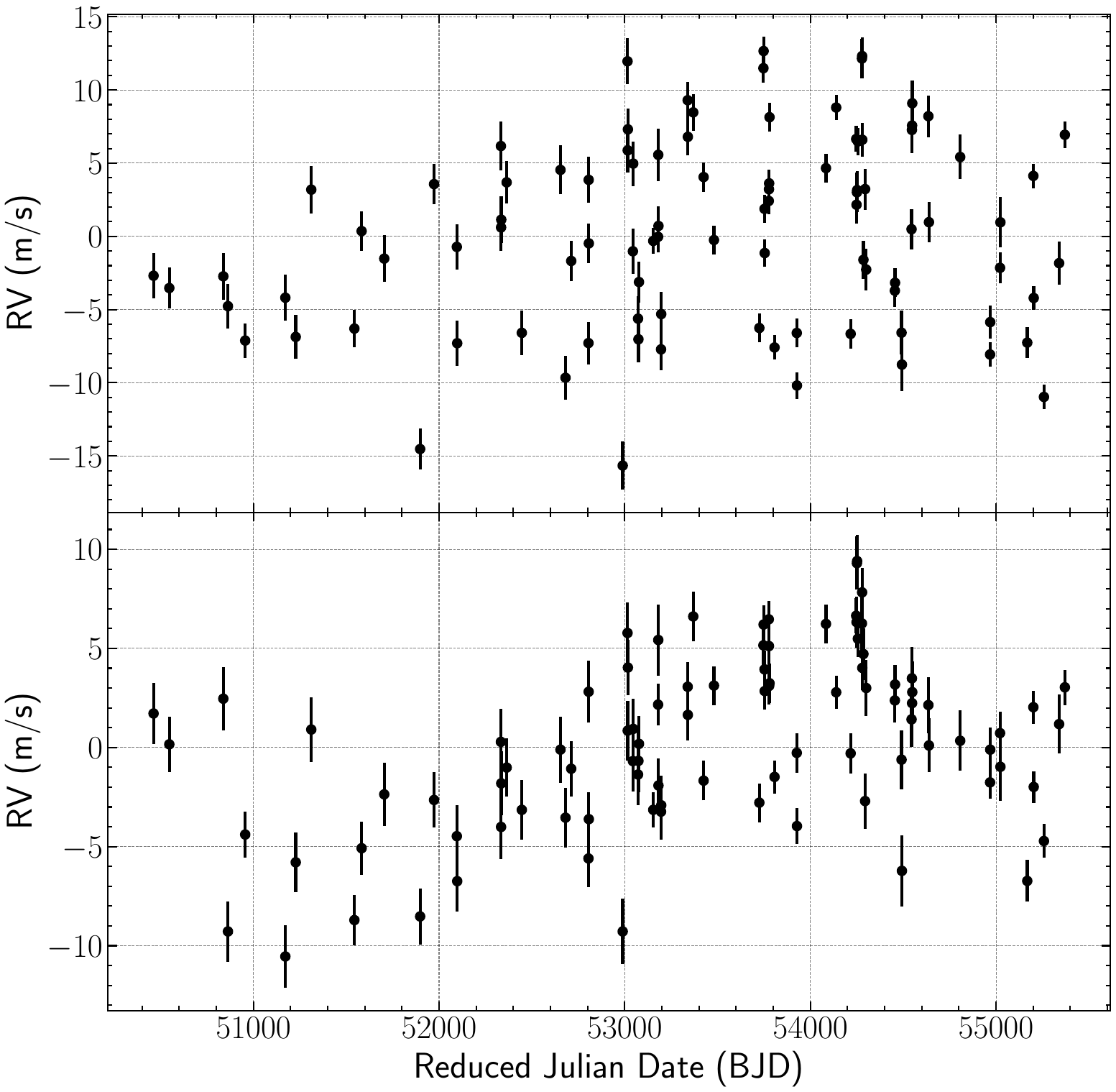}}
\caption{Radial velocity measurements of HD 99492 \citep{meschiari2011}. \textbf{Top:} Original RV measurements. \textbf{Bottom:} Residual RVs after subtracting the published orbit of planet b \citep{marcy2005}. We observe long-term variation that was first thought to be ``planet c'' but later found to be a stellar activity cycle \citep{kane2016}.}
\label{fig:hd99492_rv}
\end{figure}

\begin{figure}
\centerline{\includegraphics[width=1\linewidth]{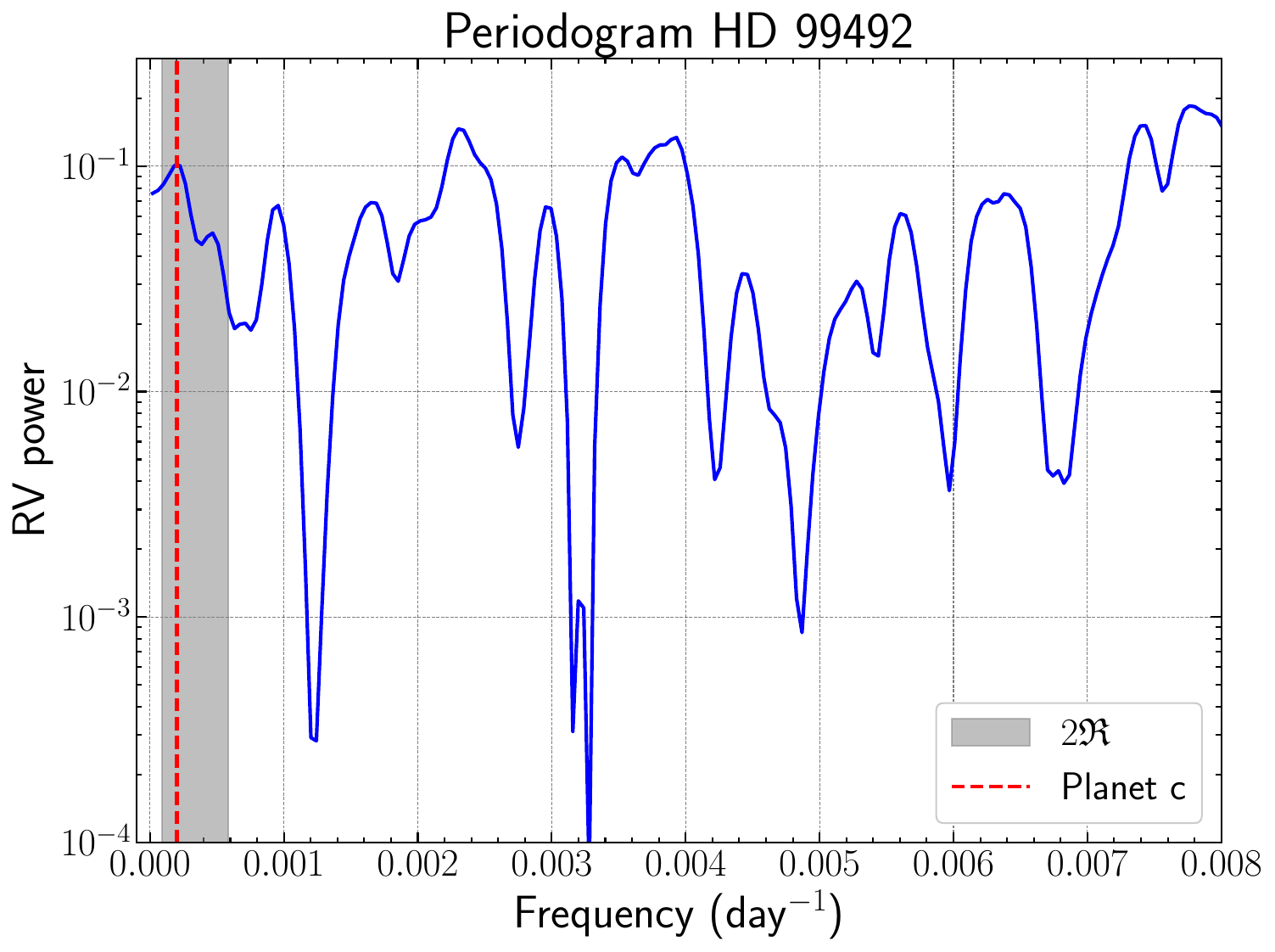}}
\caption{Periodogram of RV measurements from HD 99492, plotted in the low-frequency range. The reported frequency of planet c is highlighted with the dotted vertical line at $f = 1/4969.73$ day$^{-1}$. The frequency of the proposed planet lands inside the gray area of width $2 \mathfrak{R}$.} 
\label{fig:hd99492_per}
\end{figure}

Figure \ref{fig:hd99492_rv} shows the 93 RV measurements from \cite{meschiari2011} taken with the High Resolution Echelle Spectrometer (HIRES) at the Keck Observatory, spanning a total duration of $4908.7$ days. Figure \ref{fig:hd99492_per} shows the data's GLSP zoomed in on the low frequencies. The vertical red dotted line shows the reported frequency of the long-period planet and the gray shaded region marks the frequencies within $2 \mathfrak{R}$ of zero. The planet's frequency is $0.98 \mathfrak{R}$. The observations do not cover even one full period, leaving the ``planet'' orbit indistinguishable from a zero-frequency long-term trend. 

Long-period oscillations should be considered planet candidates instead of confirmed discoveries until the observation time baseline covers two full periods. In addition to obeying the Rayleigh criterion, such a policy would ensure that the velocity signal during each period is identical, which is true of planet orbits but not of activity cycles \citep{baliunas95}. However, strong signals from massive planets or brown dwarf companions present an exception due to their high RV amplitude. When a signal has an RV amplitude that greatly surpasses what is typically expected from stellar activity, it is almost certainly caused by an orbiting companion. For these cases, the {\it minimum} period and planet mass may be inferred, but the orbital parameters remain unconstrained until the time baseline is long enough to determine the planet's orbit.

The HD~99492 results highlight a similar story as with 55~Cnc (Sect. \ref{subsec:55Cnc}). Taking into account the Rayleigh criterion can help observers avoid constructing an inappropriate time-domain model of the long-period signals. Regardless of whether a long-period signal belongs to a planet or an activity cycle, its frequency/period should not be estimated unless $f > 2 \mathfrak{R}$. Instead, we recommend using polynomial fits to remove long-term variations \citep{cumming2008,Zechmeister2009,bonfils2013} before searching for short-term variations.

\subsection{Barnard's Star}\label{sub:barnardstar}

\begin{figure*}

\centerline{\includegraphics[width=0.9\linewidth]{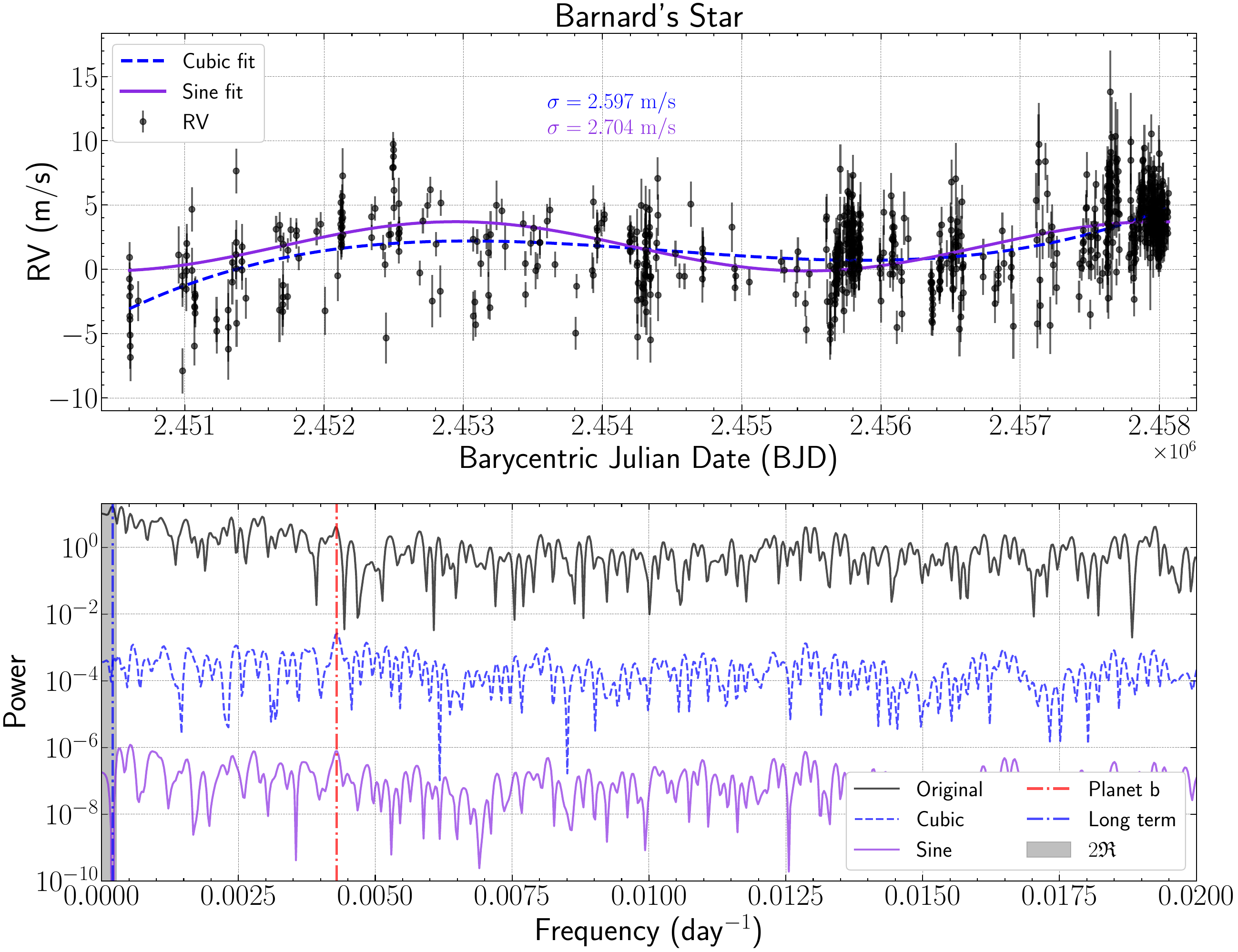}}
\caption{\textbf{Top:} Barnard's star RV data fitted with a cubic (blue dashed line) and a sinusoid (violet solid line), with values of $\sigma_r$ for each respective model. \textbf{Bottom:} GLSP of the RV data (black) and its residuals from the cubic (blue) and sine (violet) fits. 
The red dotted vertical line shows the frequency of the contested planet b and the blue dotted vertical line indicates the long-term activity cycle frequency reported by \citet{ribas2018}.}
\label{fig:barnard_star}
\end{figure*}

Barnard's star, an old M dwarf, has historically had several reports of planet detections \citep{kamp1963, kamp1969, kamp1975, kamp1982} that were ultimately shown to be false positives  \citep{gatewood1973, benedict1999, kurster2003, choi2013}. \cite{ribas2018} reported a new planet detection with a period of $232.8\pm0.4$ days. However, \cite{Lubin2021} challenged this planet detection by observing the disappearance of the peak in the GLSP when the time series was split into three epochs. They attribute the apparent 233-day signal to an alias of the activity cycle. Recently, \cite{gonzaleshernandez2024} conducted new RV observations with The Echelle SPectrograph for Rocky Exoplanets and Stable Spectroscopic Observations (ESPRESSO) \citep{espresso2021}, where they were unable to recover the $232.8\pm0.4$ day signal and instead reported a sub-Earth mass planet along with three other planet candidates.

In Sect. \ref{sec:ray_crit}, Corollary 1 states that the minimum detectable oscillation frequency in any time series is $2\mathfrak{R}$. Here we compare two different models for the star's long-term activity cycle as traced by the \citet{ribas2018} data, a cubic polynomial and a sinusoid.
Our sinusoidal model is equivalent to using a circular Keplerian ``orbit'' to subtract the long-term activity cycle from the RV, as done by Ribas et al. The cubic and the sinusoid both have four free parameters: amplitude, phase, period, vertical offset (sinusoid) and 0th to 3rd-order polynomial coefficients (cubic). Our goal is to determine whether the dataset presented by \cite{ribas2018} conclusively demonstrates that the long-term RV variations are periodic and, if so, whether it supports the reported period estimate of $\sim 6600$~day.

The top of Figure \ref{fig:barnard_star} shows the Barnard's star RVs reported by \cite{ribas2018} along with the best-fit sinusoid (purple solid line) and cubic (blue dashed line) activity models. Standard deviations $\sigma_r$ of the two fit residuals are shown in corresponding colors.
The cubic model results in a smaller value of $\sigma_r$ than the sinusoid, indicating a better fit. Our best-fit sinusoid has $P = 5000$~days, statistically indistinguishable from the \cite{ribas2018} activity-cycle period of 6600 days. The bottom panel of Figure \ref{fig:barnard_star} shows GLSPs from the original data (black), the cubic fit residual (blue dashed), and the sinusoid fit residual (purple). Gray shading shows the frequency interval in which long-period signals fail Corollary 1 (Sect. \ref{sec:ray_crit}) and vertical lines denote the activity and planet periods reported by \cite{ribas2018}. We observe that the original GLSP shows significant power at the low frequencies, but after subtracting the cubic fit most of that power disappears. The sine model also reduces some of the power at the low frequencies but there are more peaks in the residual periodogram that go above the $1\%$ FAP threshold than in the cubic fit residual. Note that neither our 5000-day best-fit sinuslid period nor the 6600-day period reported by \cite{ribas2018} pass the Rayleigh criterion.

\begin{figure*}

    \centerline{\includegraphics[width=0.8\linewidth]{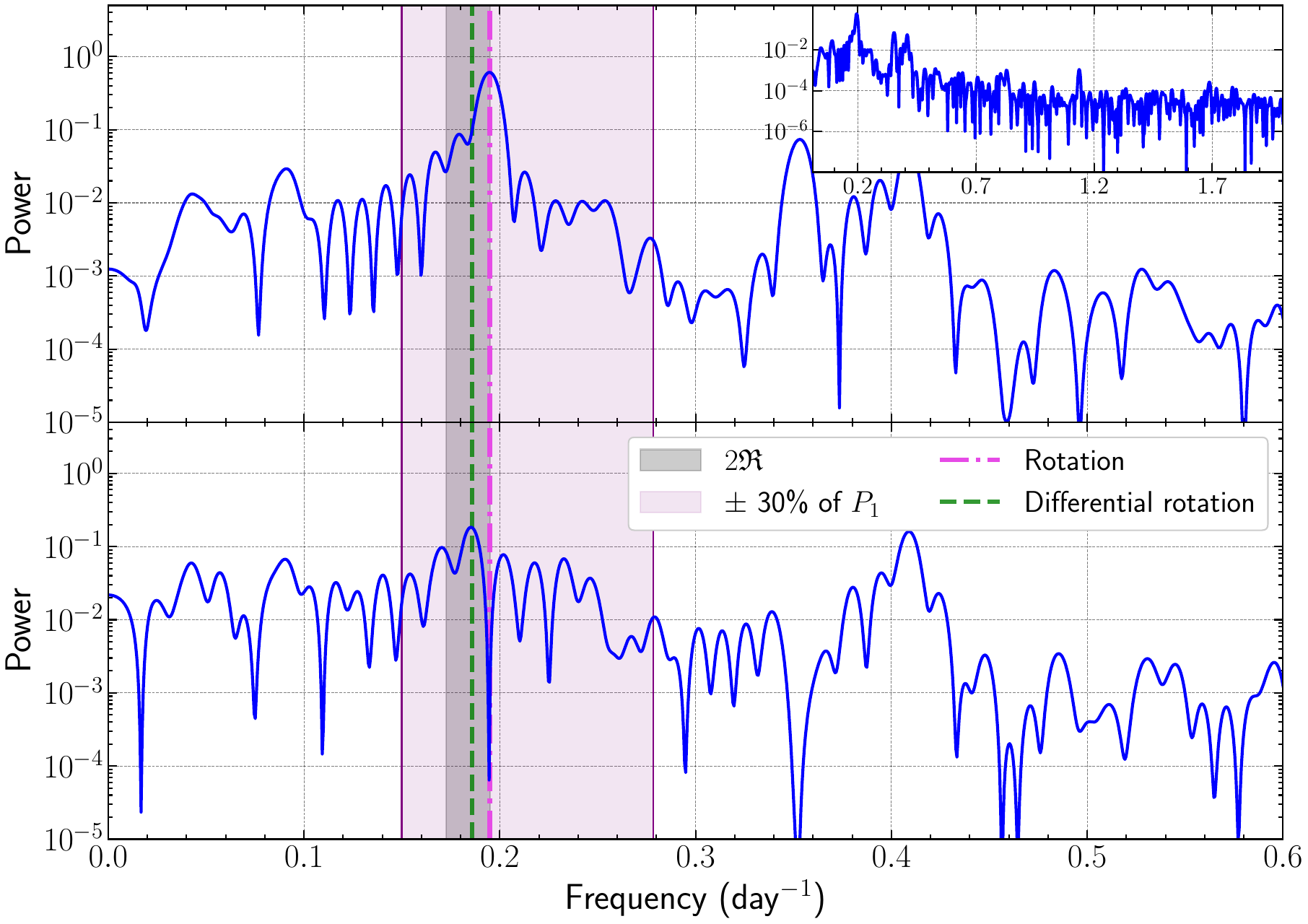}}
     \caption{\textbf{Top}: GLSP of KIC 891916's binned light curve focused on the primary rotation frequency $f_1$ (Table \ref{kepler_example}). The wide purple shaded region shows periods $P_1 \pm30\%$, and the gray shaded region has width 2$\mathfrak{R}$. The vertical lines highlight the reported rotation and differential rotation frequencies. {\bf Upper right inset:} periodogram over a wide frequency range, 0--2~day$^{-1}$. \textbf{Bottom}: GLSP of the residuals after iterating the removal of the first two most significant peaks. A peak is now visible at the frequency \citep{kepler_stars} associated with the differential rotation of this star (green vertical line).}
     \label{fig:891916_periodograms}
     
 \end{figure*}
The fact that the cubic polynomial is better at removing power that belongs in the zero-frequency periodogram bin suggests that, despite its lack of physical meaning, it is a superior detrending tool for this dataset than a sinusoid. Indeed, the recent analysis by \cite{gonzaleshernandez2024} confirms that trying to extract physical meaning from the sinusoid fit to the \citet{ribas2018} data was not successful:
augmenting the RV time series with the new ESPRESSO data reveals that the activity cycle period is $\sim 3200$ days, much shorter than the Ribas et al.\ estimate.
We emphasize that if an aperiodic function fits the star's magnetic activity better than a periodic function, it is premature to report an activity cycle period. This does not mean that the activity is not periodic,
but \emph{that the 2018 dataset did not have a sufficient time baseline for an activity period estimate}. After the time baseline increased with the addition of the ESPRESSO data, the Rayleigh criterion was satisfied for the new estimate of the activity cycle period.
Observers searching for periodic signals in new datasets need not perform the fitting experiment illustrated in Figure \ref{fig:barnard_star}; Corollary 1 of the Rayleigh criterion (Sect.\ \ref{sec:ray_crit}) is a sufficient warning of when one cannot claim a periodic signal detection. 

\begin{table*}
\centering
\caption{Primary and secondary rotation frequencies reported for two {\it Kepler} targets \citep{kepler_stars} in which the Rayleigh Criterion for distinguishing two signals is not met.} 
\begin{center}
\begin{tabular}{|c|c|c|c|c|c|c|}
\hline
\textbf{KIC} & \textbf{\textit{P\textsubscript{1}}} {[}days{]} & \textbf{\textit{f\textsubscript{1}}} {[}day\textsuperscript{-1}{]} & \textbf{\textit{P\textsubscript{2}}} {[}days{]} & \textbf{\textit{f\textsubscript{2}}} {[}day\textsuperscript{-1}{]} & \boldmath$\lvert f_1-f_2\rvert$ {[}day\textsuperscript{-1}{]} & $\mathbf{2\mathfrak{R}}$ {[}day\textsuperscript{-1}{]} \\ \hline
891916       & 5.1299                 & 0.1949                        & 5.3779                 & 0.1859                        & 0.0090                           & \multirow{2}{*}{0.0224}      \\ \cline{1-6}
1869783      & 26.1778                & 0.0382                        & 20.9422                & 0.0478                        & 0.0096                           &                              \\ \hline
\end{tabular}
\end{center}
\label{kepler_example}
\end{table*}

\subsection{{\it Kepler} differential rotators}
\label{subsec:diffrot}

In their study of stellar differential rotation and surface shear, \cite{kepler_stars} analyzed \textit{Kepler} Q3 data from a sample of 40,661 stars and reported rotation periods of 24,124 stars.
They identified differential rotation by finding the highest peak in the GLSP of each binned light curve, then fitting and subtracting out a sinusoid at that frequency. They then searched for the highest peak in a periodogram of the residuals and fitted and subtracted a second sinusoid. The iterative sinusoid fitting was performed five times to estimate the frequencies $f_i$ of a model, 
\begin{equation}
F = \sum_{i=1}^{5} A_i \sin (2 \pi f_i t + \phi_i)
\label{eq:truncated_Fourier}
\end{equation}
(where $F$ is the flux in e$^- s^{-1}$). Finally, \cite{kepler_stars} fit the model to the binned light curve, optimizing all parameters $A_i$, $f_i$, and $\phi_i$. A model with $P_i = 1/f_i$ and $|P_i - P_1| < 0.3 P_1$, where $P_1$ is the primary rotation period associated with the most significant periodogram peak (or its fundamental, if the most powerful peak is a harmonic) was assumed to indicate differential rotation.

However, stellar rotation manifests as a quasiperiodic signal instead of a purely periodic signal, meaning that (1) the apparent rotation frequency can shift over time \citep[e.g.][]{bloomfield00, yang20, dodson2022}, and (2) the amplitude and phase of any rotation signal are expected to fluctuate \citep{dumusque11, haywood14, angus18}. Amplitude and/or phase drift of the primary rotation signal or a slight inaccuracy in its frequency could create spurious periodicities in the residuals after subtracting the first Fourier component \citep{foster95, boisse11}. When the two quoted differential rotation frequencies are separated by less than one Rayleigh resolution, there is a danger that the secondary signal is not truly distinct from the primary, but is instead an artifact of the quasiperiodicity. \citet{foster95} discusses in detail how modulated sinusoids produce many closely spaced periodogram peaks. 

\subsubsection{Rayleigh Criterion Application}

Of the 24,124 stars with rotation periods reported by \citet{kepler_stars}, 18,616 stars have a second period attributed to differential rotation. The number of stars where the separation between these two frequencies is less than $2\mathfrak{R}$ is 17,081, or 91.7\% of the reported differential rotators. The number of stars where the separation is less than $\mathfrak{R}$ is also significant with a total of 8,188. Therefore, it is unclear whether the reported differential rotation periods are truly distinct from the primary rotation periods, or are artifacts of observing cadence or quasiperiodicity.

As an example, we analyze the Q3 light curves of two of the \cite{kepler_stars} targets, KIC 891916 and KIC 1869783 (Table \ref{kepler_example})\footnote{Light curves were downloaded from the MAST archive \url{https://archive.stsci.edu/kepler/data_search/search.php}. Data is available in the archive attached to this publication \cite{RRarchive}}. We followed the procedure from \cite{kepler_stars} by binning the light curves into two-hour blocks\footnote{Binning was performed with the \texttt{Lightkurve} python package \citep{lightkurve}.}. We then computed a GLSP of each binned light curve with a frequency grid oversampled by a factor of 20 (i.e.\ 20 frequency grid points per Rayleigh resolution). Table \ref{kepler_example} shows the primary and secondary rotation frequencies reported by \citet{kepler_stars}. For both stars, the separation between the reported rotation frequencies ($f_1$, $f_2$) is less than $2\mathfrak{R}$.

\begin{figure*}

    \centerline{\includegraphics[width=0.8\linewidth]{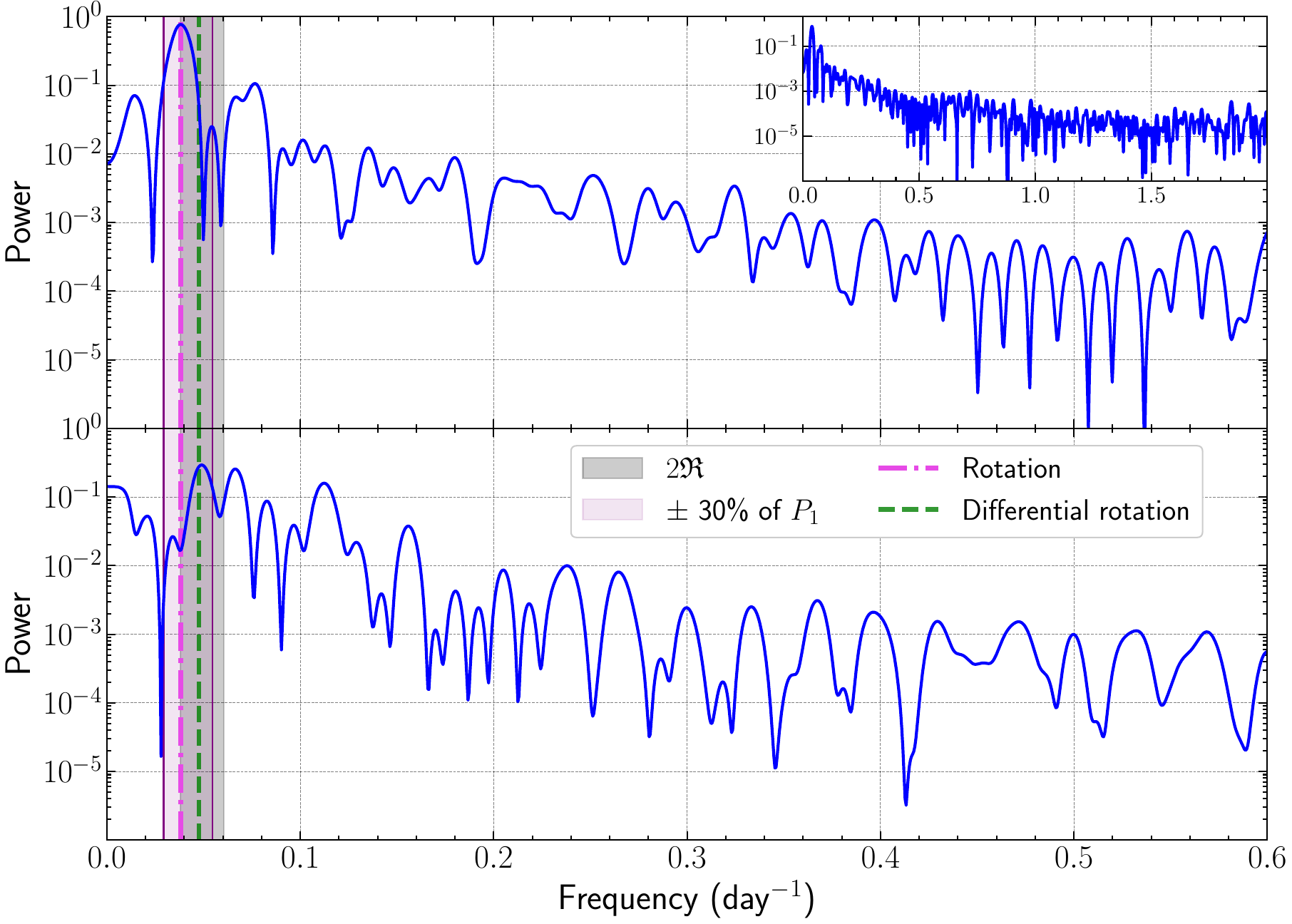}}
    \caption{\textbf{Top, upper right inset}: Same as the top panel of \ref{fig:891916_periodograms} but for KIC 1869783 \textbf{Bottom}: GLSP of its residuals after iterating the removal of the first four most significant peaks. The most significant peak is now the centered at the reported differential rotation frequency.}
    \label{fig:1869783_periodograms}
    
\end{figure*}

The top panel of Figure \ref{fig:891916_periodograms} shows a GLSP of the binned light curve of KIC~891916 zoomed in on low frequencies. The purple shaded region centered on the primary rotation frequency $f_1$ indicates the range $\pm 30\%$ $P_{1}$ over which \citet{kepler_stars} searched for secondary peaks corresponding to differential rotation, and the gray shaded region with width $2 \mathfrak{R}$ illustrates the frequency range over which signals are unresolved from $f_1$. The highly significant peak at $f_1$ is not isolated; its two slightly lower-frequency companion peaks are likely the result of modulation. The inset panel in the upper right corner, which shows the GLSP across a wider range of frequencies, reveals red noise via a downward trend in $\log \hat{S}^p(f)$. It also shows a significant harmonic of the primary rotation signal at $2 f_1$.

Following the procedure from \citet{kepler_stars}, we found the candidate oscillation at $f_2$ after performing two iterations of sinusoid fitting and subtraction\footnote{This procedure was done using the \texttt{model} method from the \texttt{LombScargle} class (\url{https://docs.astropy.org/en/stable/api/astropy.timeseries.LombScargle.html\#astropy.timeseries.LombScargle.model)}.}. The GLSP of the residuals is displayed in the bottom panel of Figure \ref{fig:891916_periodograms}. Here the highest peak aligns with the reported differential rotation period of KIC 891916 from \citet{kepler_stars}.
However, the large power density in the peak at $f_2$ results partly from the fact that it sits upon a high continuum resulting from red noise; an oscillation with similar amplitude at $f = 1.5$~day$^{-1}$ would appear to have less power in the periodogram because the noise power is lower at high frequencies. The difficulty in assessing the statistical significance of the oscillation at $f_2$ and the fact that the periodogram peak at $f_1$ is not isolated should trigger caution when reporting differential rotation. Invoking the Rayleigh criterion is a simple way to guard against spurious signal detections caused by modulation.


The top panel of Figure \ref{fig:1869783_periodograms} shows the low-frequency GLSP of the binned KIC 1869783 light curve. The bottom panel shows the GLSP of the residuals after iterative subtraction of four sinusoids, in which a peak is visible at $f_{2}$. 
Once again the log-periodogram (inset) shows red noise, which helps boost the amplitude of the peak at $f_2$ in the residual periodogram 
In fact, thanks to the high noise background at low frequencies, the peak at $f_2$ is accompanied by several neighboring peaks that also have similarly high power.

\subsubsection{Gaussian Process Model}

Here we model each {\it Kepler} light curve with a single quasiperiodic rotation signal rather than a Fourier series representing differential rotation. We use a Gaussian process \citep{haywood14, angus18} with a stochastically driven, damped simple harmonic oscillator (SHO) kernel \citep{celerite1}:
\begin{multline}
    k_\mathrm{SHO}(\tau) = S_0 \omega_0 Q e^{-\frac{\omega_0 \tau}{2Q}}\\ 
    \begin{cases}
    \cosh(\eta\omega_0\tau) + \frac{1}{2\eta Q}\sinh(\eta \omega_0  \tau), & 0<Q<1/2 \\
    2(1 +\omega_0 \tau), & Q =1/2 \\
    \cos(\eta\omega_0\tau) + \frac{1}{2\eta Q}\sin(\eta \omega_0  \tau), & 1/2<Q
    \end{cases} 
    \label{eq:SHOkernel}
\end{multline}
where $\omega_0$ is the frequency of the undamped oscillator, $Q$ is the quality factor, $S_0$ is proportional to the power spectral density at $\omega_0$ by $S(\omega_0) = \sqrt{2/\pi} S_0 Q^2$, $\tau = t_n - t_{n-1}$ represents the time lag between measurements, and $\eta = \sqrt{|1- (4Q^2)^{-1}|}$. The GP model has five free parameters instead of the 15 free parameters required for the five-term Fourier series fit of \citet{kepler_stars}.\footnote{We use Gaussian processes in order to infer physical parameters that describe rotation from the light curves. If instead the goal is to fill in the gaps in the time series, imputation via a Kalman filter would be a better option. For a review on time series imputation techniques see \cite{moritz2015comparison}.}
The free parameters are $\omega_0$, $Q$ and $S_0$, along with the time series mean and the jitter, which is an error term added to the diagonal of the covariance matrix that encompasses uncertainties not accounted for by the reported error bars. The value of $\omega_0$ is transformed into period by the relationship $\omega_0= 2\pi/P$.

\begin{figure*}
    \centerline{\includegraphics[width=1\linewidth]{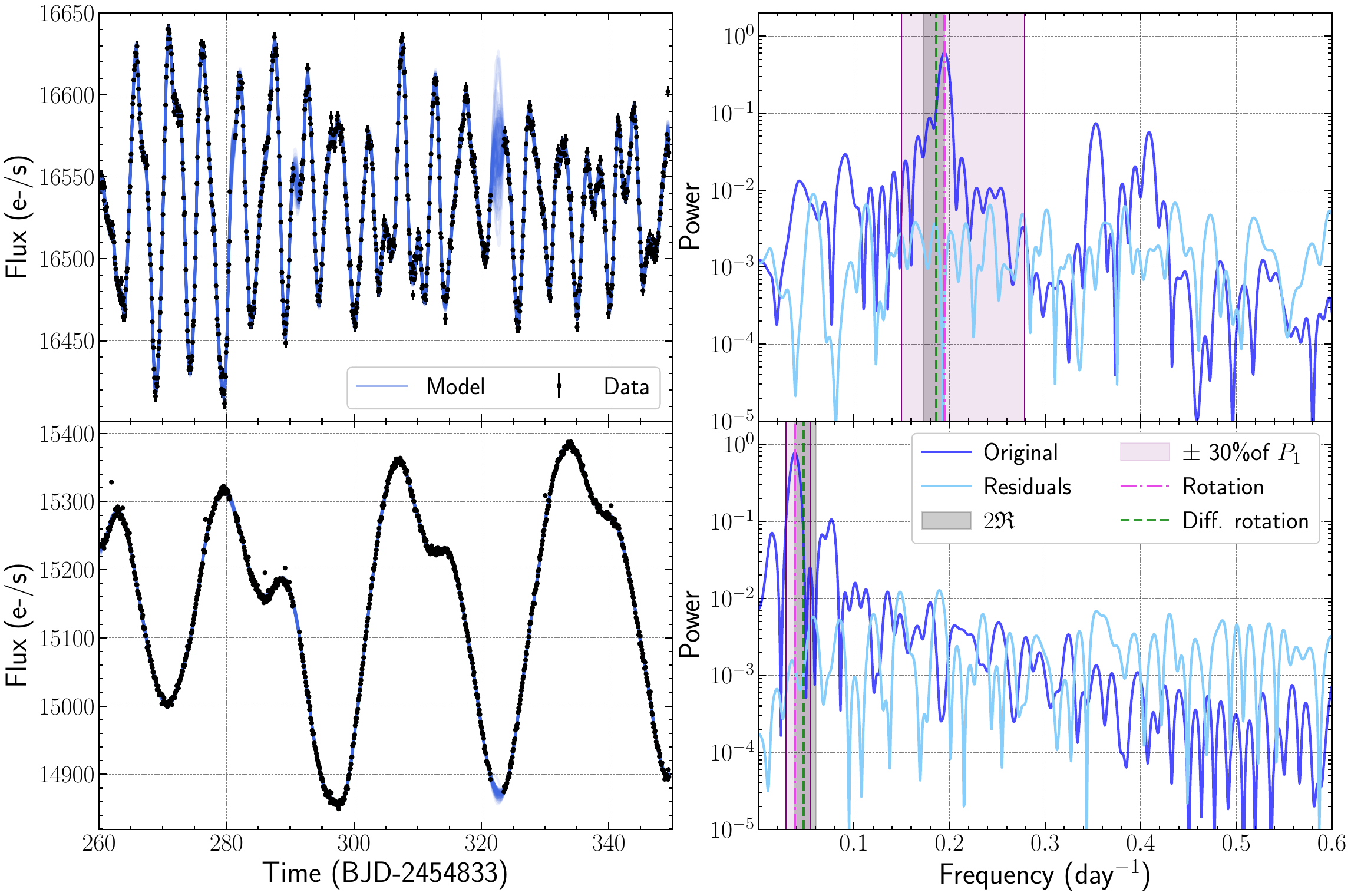}}
    \caption{{\bf Left column:} {\it Kepler} Q3 light curve of KIC 891916 (top) and  KIC 1869783 (bottom) binned into two-hour bins. The blue line represents 100 samples from the posterior distribution of the SHO model fitted to each light curve. {\bf Right column:} GLPS  of the original data and the residuals of KIC 891916 (top) and  KIC 1869783 (bottom) obtained from fitting the GP model using the median value of the parameters from the posterior distribution.}
    \label{fig:gpmodels}
    
\end{figure*}

The GP model is implemented by the \texttt{celerite2} library \citep{celerite1, celerite2}\footnote{\url{https://celerite2.readthedocs.io/en/latest/}}. We optimize the model parameters by minimizing the log-likelihood function and sample their posterior distributions using a Markov chain Monte Carlo (MCMC) method using the \texttt{PyMC} library \citep{pymc}. The left column of Figure \ref{fig:gpmodels} shows the Q3 binned light curves of each star along with 100 predictions from the posterior distribution of each GP model (blue lines). The GP faithfully reproduces the overall variation of the light curve. The right column shows the original GLSPs from Figures \ref{fig:891916_periodograms} and \ref{fig:1869783_periodograms}, along with the GLSPs of the residuals after subtracting the fit obtained from the median value of the parameters from the GP posterior distribution. Both residual periodograms show that the broad rotation peaks and their surrounding ``forests'' were successfully removed. The posterior distribution of the rotation periods have mean values of $P_1 = 21.648$ days for KIC 1869783 and $P_1 = 4.205$ days for KIC 891916. The period we extract for KIC 1869783 is within the Rayleigh resolution of the rotation period reported by \cite{kepler_stars}, while the period of KIC 891916 is separated by $3.8\mathfrak{R}$ from the literature value. This small discrepancy between the period of highest power in the GLSP and the period found by our fitting algorithm is consistent with the larger frequency width of periodogram peaks generated by quasiperiodic signals---as opposed to pure sinusoids \citep{bloomfield00}---which reduces the precision of the frequency of highest periodogram power as an estimator for the true frequency of the process \citep{angus18}. See Appendix \ref{app:cornerplots} for the complete results of the posterior distribution analysis.


Our analysis of the {\it Kepler} differential rotators demonstrates that the Rayleigh criterion is an important validation tool whenever the science goal involves detecting multiple oscillations. Given the complex ways quasiperiodic signals can manifest in periodograms, it is especially important to use caution when searching for rotation signals with small frequency separations. While we do not discourage Fourier-series models of stellar rotation, astronomers must be aware of the risks of overfitting and injecting artificial signals. 
Metrics such as the Bayesian Information Criterion, Akaike Information Criterion, and likelihood ratio test can be used to build models with the appropriate level of complexity \citep[e.g.][]{dornwallenstein19}.

By carefully taking into account the resolution of the periodogram, observers can make better decisions when choosing the time-domain model. The implications of this {\it Kepler} test case are important for upcoming space missions, such as Twinkle \citep{stotesbury22}, ARIEL \citep{ariel}, and PLATO \citep{plato2024}.

\pagebreak
\section{Conclusions} \label{sec:conclusions}

In this work, we demonstrated the importance of the Rayleigh criterion in 
period searches. 
Astronomers analyzing periodograms should ensure that the frequency separation of detected signals satisfies $\Delta f \geq 2\mathfrak{R}$ and be aware that the minimum detectable oscillation frequency is $2\mathfrak{R}$
(Corollary 1, Sect. \ref{sec:ray_crit}).
Sect. \ref{sec:syndata} highlights a handful of situations where periodogram resolution requires special attention. Artifacts from the window function can
split power from a single oscillatory process into multiple peaks in the GLSP (Sect. \ref{sub:one_sig}), but the Rayleigh criterion can be used to test whether the peaks are truly independent. The tendency towards split peaks can be avoided with the BGLS periodogram, which suppresses window function sidelobes and aliases \citep{vanderplas2018}.
In Sect.\ \ref{sub:oversample} we observed how oversampling the frequency grid does not guarantee that two independent signals can be resolved in either the GLSP or the BGLS when their frequency separation is $<2\mathfrak{R}$. For some relative phases the two signals may be resolved, but for others the single periodogram peak falls in between the two true frequencies. While one might attempt to resolve the issue by making the frequency grid denser, our analysis shows that oversampling does not improve resolution since it adds no new information to the time series (Corollary 2). In Sect. \ref{sub:var_len}, we showed that both the GLSP and the BGLS can have peaks appear at incorrect frequencies when the time baseline $T$ of a synthetic dataset does not cover two full periods of each sampled oscillation. The BGLS has the added issue that is optimized for finding a single sinusoid at a time, whereas we expect many astrophysical time series to trace multiple periodicities.

Moving on to real datasets, Sect.\ \ref{sec:results} covers several cases from the literature in which either 
two reported signals have smaller separations than $2\mathfrak{R}$ (sometimes even less than the \citet{kovacs1981} recommended resolution limit of $1.45\mathfrak{R}$), or one reported long-period signal has $f < 2 \mathfrak{R}$. In Sect.\ \ref{subsec:55Cnc}, we showed that the long-period planet 55~Cnc~d is indistinguishable from both the activity cycle and zero frequency based on the radial velocity, S-index and H$\alpha$-index. The same is true for the refuted long-period planet in the HD~99492 system (Sect. \ref{sub:hd99492}). Both of these cases highlight the importance of checking the Rayleigh criterion before constructing a time-domain model. We also demonstrated that the RV manifestation of the Barnard's star activity cycle in the \citet{ribas2018} observations is best modeled by a cubic rather than a sinusoid: the cubic removes more low-frequency power and has a lower fit RMS. The Barnard's star example emphasizes the fact that the periodicity of signals with $f < 2 \mathfrak{R}$ cannot be conclusively established or the period accurately estimated, a result confirmed by the recent analysis of \cite{gonzaleshernandez2024}. While we recognize that stellar activity is a periodic phenomenon, the dataset of \citet{ribas2018} has an insufficient time baseline for actually measuring the period---if the star is just entering or leaving a Maunder minimum state, for example, the period of the best-fit sinusoid would be misleading. For time series that appear to trace oscillations with periods similar to the time baseline, we recommend detrending with a polynomial. If an observer uses a (quasi)periodic model instead, no physical inferences should be drawn from the model.

Finally, in Sect. \ref{subsec:diffrot}, we revisited the differential rotation detections from {\it Kepler} observations of KIC~891916 and KIC~1869783, given that the primary and secondary rotation frequencies are separated by less than $2 \mathfrak{R}$. We note that this is not only the case for these two stars but for most of the \cite{kepler_stars} sample. We fitted a GP model that describes a quasiperiodic process and determined that it was effective in fitting the oscillations with a single periodicity for both stars. Observers can use metrics such as the Bayesian Information Criterion, Akaike Information Criterion, or likelihood ratio test to confirm that the chosen model is not overfitting the data by including superfluous periodic components. The Rayleigh criterion is a valuable tool for preventing false oscillation detections and mismeasured periods, and for choosing between different time-domain models of periodic phenomena.

\section*{Acknowledgments}

We thank the anonymous referee for their constructive feedback that significantly improved the manuscript. This material is based upon work supported by the University of Delaware Graduate College through the Unidel Distinguished Graduate Scholar Award. Additional funding was provided by National Science Foundation grant 2307978. Any opinions, findings, and conclusions or recommendations expressed in this material are those of the authors. This material is also based upon work supported by the U.S. Department of Energy, Office of Science, under contract number DE-AC02-06CH11357.

The submitted manuscript has been created by UChicago Argonne, LLC, Operator of Argonne National Laboratory (“Argonne”). Argonne, a U.S. Department of Energy Office of Science laboratory, is operated under Contract No. DE-AC02-06CH11357. The U.S. Government retains for itself, and others acting on its behalf, a paid-up nonexclusive, irrevocable worldwide license in said article to reproduce, prepare derivative works, distribute copies to the public, and perform publicly and display publicly, by or on behalf of the Government. The Department of Energy will provide public access to these results of federally sponsored research in accordance with the DOE Public Access Plan. http://energy.gov/downloads/doe-public-access-plan

\software{\texttt{astropy} \citep{astropy:2018}, \texttt{ArviZ} \citep{arviz}, \texttt{celerite2} \citep{celerite1,celerite2}, \texttt{Lightkurve} \citep{lightkurve}, \texttt{PyMC} \citep{pymc}.}
\bibliography{bibliography} 

\begin{thebibliography}{}
\expandafter\ifx\csname natexlab\endcsname\relax\def\natexlab#1{#1}\fi
\providecommand{\url}[1]{\href{#1}{#1}}
\providecommand{\dodoi}[1]{doi:~\href{http://doi.org/#1}{\nolinkurl{#1}}}
\providecommand{\doeprint}[1]{\href{http://ascl.net/#1}{\nolinkurl{http://ascl.net/#1}}}
\providecommand{\doarXiv}[1]{\href{https://arxiv.org/abs/#1}{\nolinkurl{https://arxiv.org/abs/#1}}}

\bibitem[{Abe \& Smith(2004)}]{Abe2004}
Abe, M., \& Smith, J.~O. 2004, Journal of The Audio Engineering Society

\bibitem[{Abril-Pla {et~al.}(2023)Abril-Pla, Andreani, Carroll, Dong, Fonnesbeck, Kochurov, Kumar, Lao, Luhmann, Martin, {et~al.}}]{pymc}
Abril-Pla, O., Andreani, V., Carroll, C., {et~al.} 2023, PeerJ Computer Science, 9, e1516

\bibitem[{{Aerts} {et~al.}(2010){Aerts}, {Christensen-Dalsgaard}, \& {Kurtz}}]{Aerts2010}
{Aerts}, C., {Christensen-Dalsgaard}, J., \& {Kurtz}, D.~W. 2010, {Asteroseismology} (Springer Science and Business Media), \dodoi{10.1007/978-1-4020-5803-5}

\bibitem[{{Anglada-Escud{\'e}} \& {Tuomi}(2012)}]{angladaescude12}
{Anglada-Escud{\'e}}, G., \& {Tuomi}, M. 2012, \aap, 548, A58, \dodoi{10.1051/0004-6361/201219910}

\bibitem[{{Anglada-Escud{\'e}} {et~al.}(2013){Anglada-Escud{\'e}}, {Tuomi}, {Gerlach}, {Barnes}, {Heller}, {Jenkins}, {Wende}, {Vogt}, {Butler}, {Reiners}, \& {Jones}}]{angladaescude13}
{Anglada-Escud{\'e}}, G., {Tuomi}, M., {Gerlach}, E., {et~al.} 2013, \aap, 556, A126, \dodoi{10.1051/0004-6361/201321331}

\bibitem[{{Angus} {et~al.}(2018){Angus}, {Morton}, {Aigrain}, {Foreman-Mackey}, \& {Rajpaul}}]{angus18}
{Angus}, R., {Morton}, T., {Aigrain}, S., {Foreman-Mackey}, D., \& {Rajpaul}, V. 2018, \mnras, 474, 2094, \dodoi{10.1093/mnras/stx2109}

\bibitem[{{Astropy Collaboration} {et~al.}(2018){Astropy Collaboration}, {Price-Whelan}, {Sip{\H{o}}cz}, {G{\"u}nther}, {Lim}, {Crawford}, {Conseil}, {Shupe}, {Craig}, {Dencheva}, {Ginsburg}, {Vand erPlas}, {Bradley}, {P{\'e}rez-Su{\'a}rez}, {de Val-Borro}, {Aldcroft}, {Cruz}, {Robitaille}, {Tollerud}, {Ardelean}, {Babej}, {Bach}, {Bachetti}, {Bakanov}, {Bamford}, {Barentsen}, {Barmby}, {Baumbach}, {Berry}, {Biscani}, {Boquien}, {Bostroem}, {Bouma}, {Brammer}, {Bray}, {Breytenbach}, {Buddelmeijer}, {Burke}, {Calderone}, {Cano Rodr{\'\i}guez}, {Cara}, {Cardoso}, {Cheedella}, {Copin}, {Corrales}, {Crichton}, {D'Avella}, {Deil}, {Depagne}, {Dietrich}, {Donath}, {Droettboom}, {Earl}, {Erben}, {Fabbro}, {Ferreira}, {Finethy}, {Fox}, {Garrison}, {Gibbons}, {Goldstein}, {Gommers}, {Greco}, {Greenfield}, {Groener}, {Grollier}, {Hagen}, {Hirst}, {Homeier}, {Horton}, {Hosseinzadeh}, {Hu}, {Hunkeler}, {Ivezi{\'c}}, {Jain}, {Jenness}, {Kanarek}, {Kendrew}, {Kern}, {Kerzendorf}, {Khvalko}, {King}, {Kirkby}, {Kulkarni},
  {Kumar}, {Lee}, {Lenz}, {Littlefair}, {Ma}, {Macleod}, {Mastropietro}, {McCully}, {Montagnac}, {Morris}, {Mueller}, {Mumford}, {Muna}, {Murphy}, {Nelson}, {Nguyen}, {Ninan}, {N{\"o}the}, {Ogaz}, {Oh}, {Parejko}, {Parley}, {Pascual}, {Patil}, {Patil}, {Plunkett}, {Prochaska}, {Rastogi}, {Reddy Janga}, {Sabater}, {Sakurikar}, {Seifert}, {Sherbert}, {Sherwood-Taylor}, {Shih}, {Sick}, {Silbiger}, {Singanamalla}, {Singer}, {Sladen}, {Sooley}, {Sornarajah}, {Streicher}, {Teuben}, {Thomas}, {Tremblay}, {Turner}, {Terr{\'o}n}, {van Kerkwijk}, {de la Vega}, {Watkins}, {Weaver}, {Whitmore}, {Woillez}, {Zabalza}, \& {Astropy Contributors}}]{astropy:2018}
{Astropy Collaboration}, {Price-Whelan}, A.~M., {Sip{\H{o}}cz}, B.~M., {et~al.} 2018, \aj, 156, 123, \dodoi{10.3847/1538-3881/aabc4f}

\bibitem[{{Baliunas} {et~al.}(1995){Baliunas}, {Donahue}, {Soon}, {Horne}, {Frazer}, {Woodard-Eklund}, {Bradford}, {Rao}, {Wilson}, {Zhang}, {Bennett}, {Briggs}, {Carroll}, {Duncan}, {Figueroa}, {Lanning}, {Misch}, {Mueller}, {Noyes}, {Poppe}, {Porter}, {Robinson}, {Russell}, {Shelton}, {Soyumer}, {Vaughan}, \& {Whitney}}]{baliunas95}
{Baliunas}, S.~L., {Donahue}, R.~A., {Soon}, W.~H., {et~al.} 1995, \apj, 438, 269, \dodoi{10.1086/175072}

\bibitem[{{Baluev}(2009)}]{baluev09}
{Baluev}, R.~V. 2009, \mnras, 395, 1541, \dodoi{10.1111/j.1365-2966.2009.14634.x}

\bibitem[{{Baluev}(2013)}]{baluev13}
---. 2013, \mnras, 429, 2052, \dodoi{10.1093/mnras/sts476}

\bibitem[{{Baluev}(2015)}]{baluev2015}
---. 2015, \mnras, 446, 1493, \dodoi{10.1093/mnras/stu2150}

\bibitem[{{Basu} \& {Antia}(2008)}]{basu08}
{Basu}, S., \& {Antia}, H.~M. 2008, \physrep, 457, 217, \dodoi{10.1016/j.physrep.2007.12.002}

\bibitem[{{Baturin} {et~al.}(2015){Baturin}, {Gorshkov}, \& {Oreshina}}]{baturin15}
{Baturin}, V.~A., {Gorshkov}, A.~B., \& {Oreshina}, A.~V. 2015, Astronomy Reports, 59, 46, \dodoi{10.1134/S1063772915010023}

\bibitem[{{Benedict} {et~al.}(1999){Benedict}, {McArthur}, {Chappell}, {Nelan}, {Jefferys}, {van Altena}, {Lee}, {Cornell}, {Shelus}, {Hemenway}, {Franz}, {Wasserman}, {Duncombe}, {Story}, {Whipple}, \& {Fredrick}}]{benedict1999}
{Benedict}, G.~F., {McArthur}, B., {Chappell}, D.~W., {et~al.} 1999, \aj, 118, 1086, \dodoi{10.1086/300975}

\bibitem[{{Black} \& {Scargle}(1982)}]{Black1982}
{Black}, D.~C., \& {Scargle}, J.~D. 1982, \apj, 263, 854, \dodoi{10.1086/160555}

\bibitem[{{Blomme} {et~al.}(2011){Blomme}, {Sarro}, {O'Donovan}, {Debosscher}, {Brown}, {Lopez}, {Dubath}, {Rimoldini}, {Charbonneau}, {Dunham}, {Mandushev}, {Ciardi}, {De Ridder}, \& {Aerts}}]{blomme11}
{Blomme}, J., {Sarro}, L.~M., {O'Donovan}, F.~T., {et~al.} 2011, \mnras, 418, 96, \dodoi{10.1111/j.1365-2966.2011.19466.x}

\bibitem[{Bloomfield(2000)}]{bloomfield00}
Bloomfield, P. 2000, Fourier analysis of time series: an introduction, 2nd ed., Wiley series in probability and statistics. Applied probability and statistics section (New York: Wiley)

\bibitem[{{Blunt} {et~al.}(2019){Blunt}, {Endl}, {Weiss}, {Cochran}, {Howard}, {MacQueen}, {Fulton}, {Henry}, {Johnson}, {Kosiarek}, {Lawson}, {Macintosh}, {Mills}, {Nielsen}, {Petigura}, {Schneider}, {Vanderburg}, {Wisniewski}, {Wittenmyer}, {Brugamyer}, {Caldwell}, {Cochran}, {Hatzes}, {Hirsch}, {Isaacson}, {Robertson}, {Roy}, \& {Shen}}]{Blunt2019}
{Blunt}, S., {Endl}, M., {Weiss}, L.~M., {et~al.} 2019, \aj, 158, 181, \dodoi{10.3847/1538-3881/ab3e63}

\bibitem[{{Boisse} {et~al.}(2011){Boisse}, {Bouchy}, {H{\'e}brard}, {Bonfils}, {Santos}, \& {Vauclair}}]{boisse11}
{Boisse}, I., {Bouchy}, F., {H{\'e}brard}, G., {et~al.} 2011, \aap, 528, A4, \dodoi{10.1051/0004-6361/201014354}

\bibitem[{{Bonfils} {et~al.}(2013){Bonfils}, {Lo Curto}, {Correia}, {Laskar}, {Udry}, {Delfosse}, {Forveille}, {Astudillo-Defru}, {Benz}, {Bouchy}, {Gillon}, {H{\'e}brard}, {Lovis}, {Mayor}, {Moutou}, {Naef}, {Neves}, {Pepe}, {Perrier}, {Queloz}, {Santos}, \& {S{\'e}gransan}}]{bonfils2013}
{Bonfils}, X., {Lo Curto}, G., {Correia}, A.~C.~M., {et~al.} 2013, \aap, 556, A110, \dodoi{10.1051/0004-6361/201220237}

\bibitem[{{Bourrier} {et~al.}(2018){Bourrier}, {Dumusque}, {Dorn}, {Henry}, {Astudillo-Defru}, {Rey}, {Benneke}, {H{\'e}brard}, {Lovis}, {Demory}, {Moutou}, \& {Ehrenreich}}]{bourrier2018}
{Bourrier}, V., {Dumusque}, X., {Dorn}, C., {et~al.} 2018, \aap, 619, A1, \dodoi{10.1051/0004-6361/201833154}

\bibitem[{{Box} \& Draper(1987)}]{box86}
{Box}, G. E.~P., \& Draper, N.~R. 1987, Empirical model-building and response surfaces (USA: John Wiley \& Sons, Inc.)

\bibitem[{Braun(2001)}]{BRAUN2001}
Braun, S. 2001, in Encyclopedia of Vibration, ed. S.~Braun (Oxford: Elsevier), 1208--1223, \dodoi{https://doi.org/10.1006/rwvb.2001.0187}

\bibitem[{{Brun} {et~al.}(2002){Brun}, {Antia}, {Chitre}, \& {Zahn}}]{brun02}
{Brun}, A.~S., {Antia}, H.~M., {Chitre}, S.~M., \& {Zahn}, J.~P. 2002, \aap, 391, 725, \dodoi{10.1051/0004-6361:20020837}

\bibitem[{{Butler} {et~al.}(1997){Butler}, {Marcy}, {Williams}, {Hauser}, \& {Shirts}}]{butler1997}
{Butler}, R.~P., {Marcy}, G.~W., {Williams}, E., {Hauser}, H., \& {Shirts}, P. 1997, \apjl, 474, L115, \dodoi{10.1086/310444}

\bibitem[{{Butler} {et~al.}(2017){Butler}, {Vogt}, {Laughlin}, {Burt}, {Rivera}, {Tuomi}, {Teske}, {Arriagada}, {Diaz}, {Holden}, \& {Keiser}}]{butler2017}
{Butler}, R.~P., {Vogt}, S.~S., {Laughlin}, G., {et~al.} 2017, \aj, 153, 208, \dodoi{10.3847/1538-3881/aa66ca}

\bibitem[{Capon(1969)}]{capon69}
Capon, J. 1969, Proceedings of the IEEE, 57, 1408, \dodoi{10.1109/PROC.1969.7278}

\bibitem[{{Chaboyer} {et~al.}(1995){Chaboyer}, {Demarque}, \& {Pinsonneault}}]{chaboyer95}
{Chaboyer}, B., {Demarque}, P., \& {Pinsonneault}, M.~H. 1995, \apj, 441, 865, \dodoi{10.1086/175408}

\bibitem[{{Choi} {et~al.}(2013){Choi}, {McCarthy}, {Marcy}, {Howard}, {Fischer}, {Johnson}, {Isaacson}, \& {Wright}}]{choi2013}
{Choi}, J., {McCarthy}, C., {Marcy}, G.~W., {et~al.} 2013, \apj, 764, 131, \dodoi{10.1088/0004-637X/764/2/131}

\bibitem[{{Christensen-Dalsgaard} \& {Gough}(1982)}]{Dalsgaard1982}
{Christensen-Dalsgaard}, J., \& {Gough}, D.~O. 1982, \mnras, 198, 141, \dodoi{10.1093/mnras/198.1.141}

\bibitem[{{Claverie} {et~al.}(1979){Claverie}, {Isaak}, {McLeod}, {van der Raay}, \& {Roca Cortes}}]{claverie79}
{Claverie}, A., {Isaak}, G.~R., {McLeod}, C.~P., {van der Raay}, H.~B., \& {Roca Cortes}, T. 1979, \nat, 282, 591, \dodoi{10.1038/282591a0}

\bibitem[{{Cumming} {et~al.}(2008){Cumming}, {Butler}, {Marcy}, {Vogt}, {Wright}, \& {Fischer}}]{cumming2008}
{Cumming}, A., {Butler}, R.~P., {Marcy}, G.~W., {et~al.} 2008, \pasp, 120, 531, \dodoi{10.1086/588487}

\bibitem[{{Dawson} \& {Fabrycky}(2010)}]{dawson2010}
{Dawson}, R.~I., \& {Fabrycky}, D.~C. 2010, \apj, 722, 937, \dodoi{10.1088/0004-637X/722/1/937}

\bibitem[{{Delisle} {et~al.}(2020){Delisle}, {Hara}, \& {S{\'e}gransan}}]{delisle20}
{Delisle}, J.~B., {Hara}, N., \& {S{\'e}gransan}, D. 2020, \aap, 635, A83, \dodoi{10.1051/0004-6361/201936905}

\bibitem[{{Dodson-Robinson} {et~al.}(2022){Dodson-Robinson}, {Delgado}, {Harrell}, \& {Haley}}]{dodson2022}
{Dodson-Robinson}, S.~E., {Delgado}, V.~R., {Harrell}, J., \& {Haley}, C.~L. 2022, \aj, 163, 169, \dodoi{10.3847/1538-3881/ac52ed}

\bibitem[{{Dorn-Wallenstein} {et~al.}(2019){Dorn-Wallenstein}, {Levesque}, \& {Davenport}}]{dornwallenstein19}
{Dorn-Wallenstein}, T.~Z., {Levesque}, E.~M., \& {Davenport}, J. R.~A. 2019, \apj, 878, 155, \dodoi{10.3847/1538-4357/ab223f}

\bibitem[{{Dumusque} {et~al.}(2011){Dumusque}, {Santos}, {Udry}, {Lovis}, \& {Bonfils}}]{dumusque11}
{Dumusque}, X., {Santos}, N.~C., {Udry}, S., {Lovis}, C., \& {Bonfils}, X. 2011, \aap, 527, A82, \dodoi{10.1051/0004-6361/201015877}

\bibitem[{{Endl} {et~al.}(2012){Endl}, {Robertson}, {Cochran}, {MacQueen}, {Brugamyer}, {Caldwell}, {Wittenmyer}, {Barnes}, \& {Gullikson}}]{endl12}
{Endl}, M., {Robertson}, P., {Cochran}, W.~D., {et~al.} 2012, \apj, 759, 19, \dodoi{10.1088/0004-637X/759/1/19}

\bibitem[{{Faria} {et~al.}(2022){Faria}, {Su{\'a}rez Mascare{\~n}o}, {Figueira}, {Silva}, {Damasso}, {Demangeon}, {Pepe}, {Santos}, {Rebolo}, {Cristiani}, {Adibekyan}, {Alibert}, {Allart}, {Barros}, {Cabral}, {D'Odorico}, {Di Marcantonio}, {Dumusque}, {Ehrenreich}, {Gonz{\'a}lez Hern{\'a}ndez}, {Hara}, {Lillo-Box}, {Lo Curto}, {Lovis}, {Martins}, {M{\'e}gevand}, {Mehner}, {Micela}, {Molaro}, {Nunes}, {Pall{\'e}}, {Poretti}, {Sousa}, {Sozzetti}, {Tabernero}, {Udry}, \& {Zapatero Osorio}}]{faria22}
{Faria}, J.~P., {Su{\'a}rez Mascare{\~n}o}, A., {Figueira}, P., {et~al.} 2022, \aap, 658, A115, \dodoi{10.1051/0004-6361/202142337}

\bibitem[{{Fischer} {et~al.}(2008){Fischer}, {Marcy}, {Butler}, {Vogt}, {Laughlin}, {Henry}, {Abouav}, {Peek}, {Wright}, {Johnson}, {McCarthy}, \& {Isaacson}}]{fischer2008}
{Fischer}, D.~A., {Marcy}, G.~W., {Butler}, R.~P., {et~al.} 2008, \apj, 675, 790, \dodoi{10.1086/525512}

\bibitem[{{Foreman-Mackey}(2018)}]{celerite2}
{Foreman-Mackey}, D. 2018, Research Notes of the American Astronomical Society, 2, 31, \dodoi{10.3847/2515-5172/aaaf6c}

\bibitem[{{Foreman-Mackey} {et~al.}(2017){Foreman-Mackey}, {Agol}, {Ambikasaran}, \& {Angus}}]{celerite1}
{Foreman-Mackey}, D., {Agol}, E., {Ambikasaran}, S., \& {Angus}, R. 2017, \aj, 154, 220, \dodoi{10.3847/1538-3881/aa9332}

\bibitem[{{Foster}(1995)}]{foster95}
{Foster}, G. 1995, \aj, 109, 1889, \dodoi{10.1086/117416}

\bibitem[{{Fuhrmeister} {et~al.}(2023){Fuhrmeister}, {Coffaro}, {Stelzer}, {Mittag}, {Czesla}, \& {Schneider}}]{fuhrmeister23}
{Fuhrmeister}, B., {Coffaro}, M., {Stelzer}, B., {et~al.} 2023, \aap, 672, A149, \dodoi{10.1051/0004-6361/202245201}

\bibitem[{{Gaia Collaboration}(2020)}]{gaia20}
{Gaia Collaboration}. 2020, VizieR Online Data Catalog, I/350, \dodoi{10.26093/cds/vizier.1350}

\bibitem[{{Gatewood} \& {Eichhorn}(1973)}]{gatewood1973}
{Gatewood}, G., \& {Eichhorn}, H. 1973, \aj, 78, 769, \dodoi{10.1086/111480}

\bibitem[{Gelman \& Rubin(1992)}]{gelman1992inference}
Gelman, A., \& Rubin, D.~B. 1992, Statistical science, 7, 457

\bibitem[{Godin(1972)}]{godin72}
Godin, G. 1972, The Analysis of Tides (University of Toronto Press)

\bibitem[{{Gonz{\'a}lez Hern{\'a}ndez} {et~al.}(2024){Gonz{\'a}lez Hern{\'a}ndez}, {Su{\'a}rez Mascare{\~n}o}, {Silva}, {Stefanov}, {Faria}, {Tabernero}, {Sozzetti}, {Rebolo}, {Pepe}, {Santos}, {Cristiani}, {Lovis}, {Dumusque}, {Figueira}, {Lillo-Box}, {Nari}, {Benatti}, {Hobson}, {Castro-Gonz{\'a}lez}, {Allart}, {Passegger}, {Zapatero Osorio}, {Adibekyan}, {Alibert}, {Allende Prieto}, {Bouchy}, {Damasso}, {D'Odorico}, {Di Marcantonio}, {Ehrenreich}, {Lo Curto}, {Santos}, {Martins}, {Mehner}, {Micela}, {Molaro}, {Nunes}, {Palle}, {Sousa}, \& {Udry}}]{gonzaleshernandez2024}
{Gonz{\'a}lez Hern{\'a}ndez}, J.~I., {Su{\'a}rez Mascare{\~n}o}, A., {Silva}, A.~M., {et~al.} 2024, \aap, 690, A79, \dodoi{10.1051/0004-6361/202451311}

\bibitem[{{Gregory}(2016)}]{gregory16}
{Gregory}, P.~C. 2016, \mnras, 458, 2604, \dodoi{10.1093/mnras/stw147}

\bibitem[{{Gregory} \& {Fischer}(2010)}]{gregory10}
{Gregory}, P.~C., \& {Fischer}, D.~A. 2010, \mnras, 403, 731, \dodoi{10.1111/j.1365-2966.2009.16233.x}

\bibitem[{{Hara} {et~al.}(2017){Hara}, {Bou{\'e}}, {Laskar}, \& {Correia}}]{hara2017}
{Hara}, N.~C., {Bou{\'e}}, G., {Laskar}, J., \& {Correia}, A.~C.~M. 2017, \mnras, 464, 1220, \dodoi{10.1093/mnras/stw2261}

\bibitem[{Harris(1978)}]{harris78}
Harris, F.~J. 1978, Proceedings of the IEEE, 66, 51

\bibitem[{{Hatzes} {et~al.}(2018){Hatzes}, {Endl}, {Cochran}, {MacQueen}, {Han}, {Lee}, {Kim}, {Mkrtichian}, {D{\"o}llinger}, {Hartmann}, {Karjalainen}, \& {Dreizler}}]{hatzes18}
{Hatzes}, A.~P., {Endl}, M., {Cochran}, W.~D., {et~al.} 2018, \aj, 155, 120, \dodoi{10.3847/1538-3881/aaa8e1}

\bibitem[{{Haywood} {et~al.}(2014){Haywood}, {Collier Cameron}, {Queloz}, {Barros}, {Deleuil}, {Fares}, {Gillon}, {Lanza}, {Lovis}, {Moutou}, {Pepe}, {Pollacco}, {Santerne}, {S{\'e}gransan}, \& {Unruh}}]{haywood14}
{Haywood}, R.~D., {Collier Cameron}, A., {Queloz}, D., {et~al.} 2014, \mnras, 443, 2517, \dodoi{10.1093/mnras/stu1320}

\bibitem[{{Horne} \& {Baliunas}(1986)}]{Horne1986}
{Horne}, J.~H., \& {Baliunas}, S.~L. 1986, \apj, 302, 757, \dodoi{10.1086/164037}

\bibitem[{{Jenkins} {et~al.}(2014){Jenkins}, {Yoma}, {Rojo}, {Mahu}, \& {Wuth}}]{jenkins14}
{Jenkins}, J.~S., {Yoma}, N.~B., {Rojo}, P., {Mahu}, R., \& {Wuth}, J. 2014, \mnras, 441, 2253, \dodoi{10.1093/mnras/stu683}

\bibitem[{{Kane} {et~al.}(2016){Kane}, {Thirumalachari}, {Henry}, {Hinkel}, {Jensen}, {Boyajian}, {Fischer}, {Howard}, {Isaacson}, \& {Wright}}]{kane2016}
{Kane}, S.~R., {Thirumalachari}, B., {Henry}, G.~W., {et~al.} 2016, \apjl, 820, L5, \dodoi{10.3847/2041-8205/820/1/L5}

\bibitem[{Kaplun {et~al.}(2023)Kaplun, Gulvanskii, Minenkov, \& Kanatov}]{kaplun23}
Kaplun, D.~I., Gulvanskii, V.~V., Minenkov, D.~V., \& Kanatov, I.~I. 2023, in 2023 Seminar on Signal Processing, 32--35, \dodoi{10.1109/IEEECONF60473.2023.10366088}

\bibitem[{{Koen}(2006)}]{Koen2006}
{Koen}, C. 2006, \mnras, 371, 1390, \dodoi{10.1111/j.1365-2966.2006.10762.x}

\bibitem[{{Kovacs}(1981)}]{kovacs1981}
{Kovacs}, G. 1981, \apss, 78, 175, \dodoi{10.1007/BF00654032}

\bibitem[{Kumar {et~al.}(2019)Kumar, Carroll, Hartikainen, \& Martin}]{arviz}
Kumar, R., Carroll, C., Hartikainen, A., \& Martin, O. 2019, Journal of Open Source Software, 4, 1143, \dodoi{10.21105/joss.01143}

\bibitem[{{K{\"u}rster} {et~al.}(2003){K{\"u}rster}, {Endl}, {Rouesnel}, {Els}, {Kaufer}, {Brillant}, {Hatzes}, {Saar}, \& {Cochran}}]{kurster2003}
{K{\"u}rster}, M., {Endl}, M., {Rouesnel}, F., {et~al.} 2003, \aap, 403, 1077, \dodoi{10.1051/0004-6361:20030396}

\bibitem[{Legendre \& Legendre(2012)}]{legendre2012numerical}
Legendre, P., \& Legendre, L. 2012, Numerical ecology (Elsevier)

\bibitem[{{Lightkurve Collaboration} {et~al.}(2018){Lightkurve Collaboration}, {Cardoso}, {Hedges}, {Gully-Santiago}, {Saunders}, {Cody}, {Barclay}, {Hall}, {Sagear}, {Turtelboom}, {Zhang}, {Tzanidakis}, {Mighell}, {Coughlin}, {Bell}, {Berta-Thompson}, {Williams}, {Dotson}, \& {Barentsen}}]{lightkurve}
{Lightkurve Collaboration}, {Cardoso}, J.~V.~d.~M., {Hedges}, C., {et~al.} 2018, {Lightkurve: Kepler and TESS time series analysis in Python}, Astrophysics Source Code Library.
\newblock \doeprint{1812.013}

\bibitem[{{Lomb}(1976)}]{lomb1976}
{Lomb}, N.~R. 1976, \apss, 39, 447, \dodoi{10.1007/BF00648343}

\bibitem[{{L{\'o}pez-Morales} {et~al.}(2014){L{\'o}pez-Morales}, {Triaud}, {Rodler}, {Dumusque}, {Buchhave}, {Harutyunyan}, {Hoyer}, {Alonso}, {Gillon}, {Kaib}, {Latham}, {Lovis}, {Pepe}, {Queloz}, {Raymond}, {S{\'e}gransan}, {Waldmann}, \& {Udry}}]{lopezmorales14}
{L{\'o}pez-Morales}, M., {Triaud}, A. H.~M.~J., {Rodler}, F., {et~al.} 2014, \apjl, 792, L31, \dodoi{10.1088/2041-8205/792/2/L31}

\bibitem[{{Loumos} \& {Deeming}(1978)}]{Loumos1978}
{Loumos}, G.~L., \& {Deeming}, T.~J. 1978, \apss, 56, 285, \dodoi{10.1007/BF01879560}

\bibitem[{{Lubin} {et~al.}(2021){Lubin}, {Robertson}, {Stefansson}, {Ninan}, {Mahadevan}, {Endl}, {Ford}, {Wright}, {Beard}, {Bender}, {Cochran}, {Diddams}, {Fredrick}, {Halverson}, {Kanodia}, {Metcalf}, {Ramsey}, {Roy}, {Schwab}, \& {Terrien}}]{Lubin2021}
{Lubin}, J., {Robertson}, P., {Stefansson}, G., {et~al.} 2021, \aj, 162, 61, \dodoi{10.3847/1538-3881/ac0057}

\bibitem[{Lyard {et~al.}(2021)Lyard, Allain, Cancet, Carrère, \& Picot}]{fes2014}
Lyard, F.~H., Allain, D.~J., Cancet, M., Carrère, L., \& Picot, N. 2021, Ocean Science, 17, 615, \dodoi{10.5194/os-17-615-2021}

\bibitem[{{Marcy} {et~al.}(2002){Marcy}, {Butler}, {Fischer}, {Laughlin}, {Vogt}, {Henry}, \& {Pourbaix}}]{marcy2002}
{Marcy}, G.~W., {Butler}, R.~P., {Fischer}, D.~A., {et~al.} 2002, \apj, 581, 1375, \dodoi{10.1086/344298}

\bibitem[{{Marcy} {et~al.}(2005){Marcy}, {Butler}, {Vogt}, {Fischer}, {Henry}, {Laughlin}, {Wright}, \& {Johnson}}]{marcy2005}
{Marcy}, G.~W., {Butler}, R.~P., {Vogt}, S.~S., {et~al.} 2005, \apj, 619, 570, \dodoi{10.1086/426384}

\bibitem[{{McArthur} {et~al.}(2004){McArthur}, {Endl}, {Cochran}, {Benedict}, {Fischer}, {Marcy}, {Butler}, {Naef}, {Mayor}, {Queloz}, {Udry}, \& {Harrison}}]{macarthut2004}
{McArthur}, B.~E., {Endl}, M., {Cochran}, W.~D., {et~al.} 2004, \apjl, 614, L81, \dodoi{10.1086/425561}

\bibitem[{{Meschiari} {et~al.}(2011){Meschiari}, {Laughlin}, {Vogt}, {Butler}, {Rivera}, {Haghighipour}, \& {Jalowiczor}}]{meschiari2011}
{Meschiari}, S., {Laughlin}, G., {Vogt}, S.~S., {et~al.} 2011, \apj, 727, 117, \dodoi{10.1088/0004-637X/727/2/117}

\bibitem[{Montgomery \& O'donoghue(1999)}]{montgomery1999derivation}
Montgomery, M., \& O'donoghue, D. 1999, Delta Scuti Star Newsletter, vol. 13, p. 28, 13, 28

\bibitem[{Moritz {et~al.}(2015)Moritz, Sard{\'a}, Bartz-Beielstein, Zaefferer, \& Stork}]{moritz2015comparison}
Moritz, S., Sard{\'a}, A., Bartz-Beielstein, T., Zaefferer, M., \& Stork, J. 2015, arXiv preprint arXiv:1510.03924

\bibitem[{{Mortier} \& {Collier Cameron}(2017)}]{mortier2017}
{Mortier}, A., \& {Collier Cameron}, A. 2017, \aap, 601, A110, \dodoi{10.1051/0004-6361/201630201}

\bibitem[{{Mortier} {et~al.}(2015){Mortier}, {Faria}, {Correia}, {Santerne}, \& {Santos}}]{mortier15}
{Mortier}, A., {Faria}, J.~P., {Correia}, C.~M., {Santerne}, A., \& {Santos}, N.~C. 2015, \aap, 573, A101, \dodoi{10.1051/0004-6361/201424908}

\bibitem[{{Naef} {et~al.}(2004){Naef}, {Mayor}, {Beuzit}, {Perrier}, {Queloz}, {Sivan}, \& {Udry}}]{naef2004}
{Naef}, D., {Mayor}, M., {Beuzit}, J.~L., {et~al.} 2004, \aap, 414, 351, \dodoi{10.1051/0004-6361:20034091}

\bibitem[{Naidu(1995)}]{naidu95}
Naidu, P. 1995, Modern spectrum analysis of time series (CRC Press, Boca Raton)

\bibitem[{{Nelson} {et~al.}(2014){Nelson}, {Ford}, {Wright}, {Fischer}, {von Braun}, {Howard}, {Payne}, \& {Dindar}}]{nelson2014}
{Nelson}, B.~E., {Ford}, E.~B., {Wright}, J.~T., {et~al.} 2014, \mnras, 441, 442, \dodoi{10.1093/mnras/stu450}

\bibitem[{{O'Toole} {et~al.}(2009){O'Toole}, {Tinney}, {Jones}, {Butler}, {Marcy}, {Carter}, \& {Bailey}}]{otoole09}
{O'Toole}, S.~J., {Tinney}, C.~G., {Jones}, H.~R.~A., {et~al.} 2009, \mnras, 392, 641, \dodoi{10.1111/j.1365-2966.2008.14051.x}

\bibitem[{{Pan} {et~al.}(2020){Pan}, {Fu}, {Zong}, {Zhang}, {Wang}, \& {Li}}]{yang20}
{Pan}, Y., {Fu}, J.-N., {Zong}, W., {et~al.} 2020, \apj, 905, 67, \dodoi{10.3847/1538-4357/abc250}

\bibitem[{{Pepe} {et~al.}(2021){Pepe}, {Cristiani}, {Rebolo}, {Santos}, {Dekker}, {Cabral}, {Di Marcantonio}, {Figueira}, {Lo Curto}, {Lovis}, {Mayor}, {M{\'e}gevand}, {Molaro}, {Riva}, {Zapatero Osorio}, {Amate}, {Manescau}, {Pasquini}, {Zerbi}, {Adibekyan}, {Abreu}, {Affolter}, {Alibert}, {Aliverti}, {Allart}, {Allende Prieto}, {{\'A}lvarez}, {Alves}, {Avila}, {Baldini}, {Bandy}, {Barros}, {Benz}, {Bianco}, {Borsa}, {Bourrier}, {Bouchy}, {Broeg}, {Calderone}, {Cirami}, {Coelho}, {Conconi}, {Coretti}, {Cumani}, {Cupani}, {D'Odorico}, {Damasso}, {Deiries}, {Delabre}, {Demangeon}, {Dumusque}, {Ehrenreich}, {Faria}, {Fragoso}, {Genolet}, {Genoni}, {G{\'e}nova Santos}, {Gonz{\'a}lez Hern{\'a}ndez}, {Hughes}, {Iwert}, {Kerber}, {Knudstrup}, {Landoni}, {Lavie}, {Lillo-Box}, {Lizon}, {Maire}, {Martins}, {Mehner}, {Micela}, {Modigliani}, {Monteiro}, {Monteiro}, {Moschetti}, {Murphy}, {Nunes}, {Oggioni}, {Oliveira}, {Oshagh}, {Pall{\'e}}, {Pariani}, {Poretti}, {Rasilla}, {Rebord{\~a}o}, {Redaelli}, {Santana Tschudi},
  {Santin}, {Santos}, {S{\'e}gransan}, {Schmidt}, {Segovia}, {Sosnowska}, {Sozzetti}, {Sousa}, {Span{\`o}}, {Su{\'a}rez Mascare{\~n}o}, {Tabernero}, {Tenegi}, {Udry}, \& {Zanutta}}]{espresso2021}
{Pepe}, F., {Cristiani}, S., {Rebolo}, R., {et~al.} 2021, \aap, 645, A96, \dodoi{10.1051/0004-6361/202038306}

\bibitem[{{Press} {et~al.}(1987){Press}, {Flannery}, {Teukolsky}, {Vetterling}, \& {Gould}}]{Press1987}
{Press}, W.~H., {Flannery}, B.~P., {Teukolsky}, S.~A., {Vetterling}, W.~T., \& {Gould}, H. 1987, American Journal of Physics, 55, 90, \dodoi{10.1119/1.14981}

\bibitem[{{Rajpaul} {et~al.}(2015){Rajpaul}, {Aigrain}, {Osborne}, {Reece}, \& {Roberts}}]{rajpaul15}
{Rajpaul}, V., {Aigrain}, S., {Osborne}, M.~A., {Reece}, S., \& {Roberts}, S. 2015, \mnras, 452, 2269, \dodoi{10.1093/mnras/stv1428}

\bibitem[{Ramirez~Delgado(2023)}]{RRarchive}
Ramirez~Delgado, V. 2023, The Rayleigh Criterion: Resolution Limits of Astronomical Periodograms,  Zenodo, \dodoi{10.5281/ZENODO.10039084}

\bibitem[{{Rauer} {et~al.}(2024){Rauer}, {Aerts}, {Cabrera}, {Deleuil}, {Erikson}, {Gizon}, {Goupil}, {Heras}, {Lorenzo-Alvarez}, {Marliani}, {Martin-Garcia}, {Mas-Hesse}, {O'Rourke}, {Osborn}, {Pagano}, {Piotto}, {Pollacco}, {Ragazzoni}, {Ramsay}, {Udry}, {Appourchaux}, {Benz}, {Brandeker}, {G{\"u}del}, {Janot-Pacheco}, {Kabath}, {Kjeldsen}, {Min}, {Santos}, {Smith}, {Suarez}, {Werner}, {Aboudan}, {Abreu}, {Acu a}, {Adams}, {Adibekyan}, {Affer}, {Agneray}, {Agnor}, {Aguirre B{\o}rsen-Koch}, {Ahmed}, {Aigrain}, {Al-Bahlawan}, {Alcacera Gil}, {Alei}, {Alencar}, {Alexander}, {Alfonso-Garz{\'o}n}, {Alibert}, {Allende Prieto}, {Almeida}, {Alonso Sobrino}, {Altavilla}, {Althaus}, {Alonso Alvarez Trujillo}, {Amarsi}, {Ammler-von Eiff}, {Am{\^o}res}, {Andrade}, {Antoniadis-Karnavas}, {Ant{\'o}nio}, {Aparicio del Moral}, {Appolloni}, {Arena}, {Armstrong}, {Aroca Aliaga}, {Asplund}, {Audenaert}, {Auricchio}, {Avelino}, {Baeke}, {Bailli{\'e}}, {Balado}, {Ballber Balaguer{\'o}}, {Balestra}, {Ball}, {Ballans}, {Ballot},
  {Barban}, {Barbary}, {Barbieri}, {Barcel{\'o} Forteza}, {Barker}, {Barklem}, {Barnes}, {Barrado Navascues}, {Barragan}, {Baruteau}, {Basu}, {Baudin}, {Baumeister}, {Bayliss}, {Bazot}, {Beck}, {Bedding}, {Belkacem}, {Bellinger}, {Benatti}, {Benomar}, {B{\'e}rard}, {Bergemann}, {Bergomi}, {Bernardo}, {Biazzo}, {Bignamini}, {Bigot}, {Billot}, {Binet}, {Biondi}, {Biondi}, {Birch}, {Bitsch}, {Bluhm Ceballos}, {B{\'o}di}, {Bogn{\'a}r}, {Boisse}, {Bolmont}, {Bonanno}, {Bonavita}, {Bonfanti}, {Bonfils}, {Bonito}, {Bonomo}, {B{\"o}rner}, {Boro Saikia}, {Borreguero Mart{\'\i}n}, {Borsa}, {Borsato}, {Bossini}, {Bouchy}, {Bou{\'e}}, {Boufleur}, {Boumier}, {Bourrier}, {Bowman}, {Bozzo}, {Bradley}, {Bray}, {Bressan}, {Breton}, {Brienza}, {Brito}, {Brogi}, {Brown}, {Brown}, {Brun}, {Bruno}, {Bruns}, {Buchhave}, {Bugnet}, {Buldgen}, {Burgess}, {Busatta}, {Busso}, {Buzasi}, {Caballero}, {Cabral}, {Cabrero Gomez}, {Calderone}, {Cameron}, {Cameron}, {Campante}, {Campos Gestal}, {Canto Martins}, {Cara}, {Carone}, {Carrasco},
  {Casagrande}, {Casewell}, {Cassisi}, {Castellani}, {Castro}, {Catala}, {Catal{\'a}n Fern{\'a}ndez}, {Catelan}, {Cegla}, {Cerruti}, {Cessa}, {Chadid}, {Chaplin}, {Charpinet}, {Chiappini}, {Chiarucci}, {Chiavassa}, {Chinellato}, {Chirulli}, {Christensen-Dalsgaard}, {Church}, {Claret}, {Clarke}, {Claudi}, {Clermont}, {Coelho}, {Coelho}, {Cogato}, {Colom{\'e}}, {Condamin}, {Conde Garc{\'\i}a}, \& {Conseil}}]{plato2024}
{Rauer}, H., {Aerts}, C., {Cabrera}, J., {et~al.} 2024, arXiv e-prints, arXiv:2406.05447, \dodoi{10.48550/arXiv.2406.05447}

\bibitem[{Rayleigh(1879)}]{rayleigh1879}
Rayleigh, L. 1879, The London, Edinburgh, and Dublin Philosophical Magazine and Journal of Science, 8, 261

\bibitem[{{Reinhold} \& {Gizon}(2015)}]{reinhold15}
{Reinhold}, T., \& {Gizon}, L. 2015, \aap, 583, A65, \dodoi{10.1051/0004-6361/201526216}

\bibitem[{Reinhold {et~al.}(2013)Reinhold, Reiners, \& Basri}]{kepler_stars}
Reinhold, T., Reiners, A., \& Basri, G. 2013, Astronomy \& Astrophysics, 560, A4, \dodoi{10.1051/0004-6361/201321970}

\bibitem[{{Ribas} {et~al.}(2018){Ribas}, {Tuomi}, {Reiners}, {Butler}, {Morales}, {Perger}, {Dreizler}, {Rodr{\'\i}guez-L{\'o}pez}, {Gonz{\'a}lez Hern{\'a}ndez}, {Rosich}, {Feng}, {Trifonov}, {Vogt}, {Caballero}, {Hatzes}, {Herrero}, {Jeffers}, {Lafarga}, {Murgas}, {Nelson}, {Rodr{\'\i}guez}, {Strachan}, {Tal-Or}, {Teske}, {Toledo-Padr{\'o}n}, {Zechmeister}, {Quirrenbach}, {Amado}, {Azzaro}, {B{\'e}jar}, {Barnes}, {Berdi{\~n}as}, {Burt}, {Coleman}, {Cort{\'e}s-Contreras}, {Crane}, {Engle}, {Guinan}, {Haswell}, {Henning}, {Holden}, {Jenkins}, {Jones}, {Kaminski}, {Kiraga}, {K{\"u}rster}, {Lee}, {L{\'o}pez-Gonz{\'a}lez}, {Montes}, {Morin}, {Ofir}, {Pall{\'e}}, {Rebolo}, {Reffert}, {Schweitzer}, {Seifert}, {Shectman}, {Staab}, {Street}, {Su{\'a}rez Mascare{\~n}o}, {Tsapras}, {Wang}, \& {Anglada-Escud{\'e}}}]{ribas2018}
{Ribas}, I., {Tuomi}, M., {Reiners}, A., {et~al.} 2018, \nat, 563, 365, \dodoi{10.1038/s41586-018-0677-y}

\bibitem[{{Rickman} {et~al.}(2019){Rickman}, {S{\'e}gransan}, {Marmier}, {Udry}, {Bouchy}, {Lovis}, {Mayor}, {Pepe}, {Queloz}, {Santos}, {Allart}, {Bonvin}, {Bratschi}, {Cersullo}, {Chazelas}, {Choplin}, {Conod}, {Deline}, {Delisle}, {Dos Santos}, {Figueira}, {Giles}, {Girard}, {Lavie}, {Martin}, {Motalebi}, {Nielsen}, {Osborn}, {Ottoni}, {Raimbault}, {Rey}, {Roger}, {Seidel}, {Stalport}, {Su{\'a}rez Mascare{\~n}o}, {Triaud}, {Turner}, {Weber}, \& {Wyttenbach}}]{Rickman2019}
{Rickman}, E.~L., {S{\'e}gransan}, D., {Marmier}, M., {et~al.} 2019, \aap, 625, A71, \dodoi{10.1051/0004-6361/201935356}

\bibitem[{Rosenthal {et~al.}(2021)Rosenthal, Fulton, Hirsch, Isaacson, Howard, Dedrick, Sherstyuk, Blunt, Petigura, Knutson, Behmard, Chontos, Crepp, Crossfield, Dalba, Fischer, Henry, Kane, Kosiarek, Marcy, Rubenzahl, Weiss, \& Wright}]{Rosenthal2021}
Rosenthal, L.~J., Fulton, B.~J., Hirsch, L.~A., {et~al.} 2021, The Astrophysical Journal Supplement Series, 255, 8, \dodoi{10.3847/1538-4365/abe23c}

\bibitem[{{Scargle}(1982)}]{scargle1982}
{Scargle}, J.~D. 1982, \apj, 263, 835, \dodoi{10.1086/160554}

\bibitem[{Schuster(1898)}]{schuster1898}
Schuster, A. 1898, Terrestrial Magnetism, 3, 13

\bibitem[{{Schwarzenberg-Czerny}(1991)}]{Schwarzenberg-Czerny1991}
{Schwarzenberg-Czerny}, A. 1991, \mnras, 253, 198, \dodoi{10.1093/mnras/253.2.198}

\bibitem[{Shumway \& Stoffer(2001)}]{shumwaystoffer}
Shumway, R.~H., \& Stoffer, D.~S. 2001, Time Series Analysis and Its Applications, 4th edn. (New York: Springer)

\bibitem[{Smith(2011)}]{smith2011spectral}
Smith, J.~O. 2011, Spectral audio signal processing (W3K)

\bibitem[{{Stalport} {et~al.}(2023){Stalport}, {Cretignier}, {Udry}, {John}, {Wilson}, {Delisle}, {Bonomo}, {Buchhave}, {Charbonneau}, {Dalal}, {Damasso}, {Di Fabrizio}, {Dumusque}, {Fiorenzano}, {Harutyunyan}, {Haywood}, {Latham}, {L{\'o}pez-Morales}, {Lorenzi}, {Lovis}, {Malavolta}, {Molinari}, {Mortier}, {Pedani}, {Pepe}, {Pinamonti}, {Poretti}, {Rice}, \& {Sozzetti}}]{stalport2023}
{Stalport}, M., {Cretignier}, M., {Udry}, S., {et~al.} 2023, \aap, 678, A90, \dodoi{10.1051/0004-6361/202346887}

\bibitem[{{Stotesbury} {et~al.}(2022){Stotesbury}, {Edwards}, {Lavigne}, {Pesquita}, {Veilleux}, {Windred}, {Al-Refaie}, {Bradley}, {Ma}, {Savini}, {Tinetti}, {Birnstiel}, {Dodson-Robinson}, {Ercolano}, {Feliz}, {Hernitschek}, {Holdsworth}, {Jiang}, {Griffin}, {Lowson}, {Molaverdikhani}, {Neilson}, {Phillips}, {Preibisch}, {Sarkar}, {Stassun}, {Ward-Thompson}, {Wright}, {Yang}, {Yeh}, {Zhou}, {Archer}, {Barrathwaj Raman Mohan}, {Joshua}, {Tessenyi}, {Tennyson}, \& {Wilcock}}]{stotesbury22}
{Stotesbury}, I., {Edwards}, B., {Lavigne}, J.-F., {et~al.} 2022, in Society of Photo-Optical Instrumentation Engineers (SPIE) Conference Series, Vol. 12180, Space Telescopes and Instrumentation 2022: Optical, Infrared, and Millimeter Wave, ed. L.~E. {Coyle}, S.~{Matsuura}, \& M.~D. {Perrin}, 1218033, \dodoi{10.1117/12.2641373}

\bibitem[{{Su{\'a}rez Mascare{\~n}o} {et~al.}(2023){Su{\'a}rez Mascare{\~n}o}, {Gonz{\'a}lez-{\'A}lvarez}, {Zapatero Osorio}, {Lillo-Box}, {Faria}, {Passegger}, {Gonz{\'a}lez Hern{\'a}ndez}, {Figueira}, {Sozzetti}, {Rebolo}, {Pepe}, {Santos}, {Cristiani}, {Lovis}, {Silva}, {Ribas}, {Amado}, {Caballero}, {Quirrenbach}, {Reiners}, {Zechmeister}, {Adibekyan}, {Alibert}, {B{\'e}jar}, {Benatti}, {D'Odorico}, {Damasso}, {Delisle}, {Di Marcantonio}, {Dreizler}, {Ehrenreich}, {Hatzes}, {Hara}, {Henning}, {Kaminski}, {L{\'o}pez-Gonz{\'a}lez}, {Martins}, {Micela}, {Montes}, {Pall{\'e}}, {Pedraz}, {Rodr{\'\i}guez}, {Rodr{\'\i}guez-L{\'o}pez}, {Tal-Or}, {Sousa}, \& {Udry}}]{suarezmascareno2023}
{Su{\'a}rez Mascare{\~n}o}, A., {Gonz{\'a}lez-{\'A}lvarez}, E., {Zapatero Osorio}, M.~R., {et~al.} 2023, \aap, 670, A5, \dodoi{10.1051/0004-6361/202244991}

\bibitem[{Thomson {et~al.}(2007)Thomson, Lanzerotti, Vernon, Lessard, \& Smith}]{Thomson2007SolarMS}
Thomson, D.~J., Lanzerotti, L.~J., Vernon, F.~L., Lessard, M.~R., \& Smith, L. T.~P. 2007, Proceedings of the IEEE, 95, 1085

\bibitem[{Thomson \& Emery(2014)}]{thomson2014}
Thomson, R.~E., \& Emery, W.~J. 2014 (Elsevier), \dodoi{10.1016/C2010-0-66362-0}

\bibitem[{Tinetti {et~al.}(2018)Tinetti, Drossart, Eccleston, Hartogh, Heske, Leconte, Micela, Ollivier, Pilbratt, Puig, Turrini, Vandenbussche, Wolkenberg, Beaulieu, Buchave, Ferus, Griffin, Guedel, Justtanont, Lagage, Machado, Malaguti, Min, N{\o}rgaard-Nielsen, Rataj, Ray, Ribas, Swain, Szabo, Werner, Barstow, Burleigh, Cho, du~Foresto, Coustenis, Decin, Encrenaz, Galand, Gillon, Helled, Morales, Mu{\~n}oz, Moneti, Pagano, Pascale, Piccioni, Pinfield, Sarkar, Selsis, Tennyson, Triaud, Venot, Waldmann, Waltham, Wright, Amiaux, Augu{\`e}res, Berth{\'e}, Bezawada, Bishop, Bowles, Coffey, Colom{\'e}, Crook, Crouzet, Da~Peppo, Sanz, Focardi, Frericks, Hunt, Kohley, Middleton, Morgante, Ottensamer, Pace, Pearson, Stamper, Symonds, Rengel, Renotte, Ade, Affer, Alard, Allard, Altieri, Andr{\'e}, Arena, Argyriou, Aylward, Baccani, Bakos, Banaszkiewicz, Barlow, Batista, Bellucci, Benatti, Bernardi, B{\'e}zard, Blecka, Bolmont, Bonfond, Bonito, Bonomo, Brucato, Brun, Bryson, Bujwan, Casewell, Charnay, Pestellini,
  Chen, Ciaravella, Claudi, Cl{\'e}dassou, Damasso, Damiano, Danielski, Deroo, Di~Giorgio, Dominik, Doublier, Doyle, Doyon, Drummond, Duong, Eales, Edwards, Farina, Flaccomio, Fletcher, Forget, Fossey, Fr{\"a}nz, Fujii, Garc{\'\i}a-Piquer, Gear, Geoffray, G{\'e}rard, Gesa, Gomez, Graczyk, Griffith, Grodent, Guarcello, Gustin, Hamano, Hargrave, Hello, Heng, Herrero, Hornstrup, Hubert, Ida, Ikoma, Iro, Irwin, Jarchow, Jaubert, Jones, Julien, Kameda, Kerschbaum, Kervella, Koskinen, Krijger, Krupp, Lafarga, Landini, Lellouch, Leto, Luntzer, Rank-L{\"u}ftinger, Maggio, Maldonado, Maillard, Mall, Marquette, Mathis, Maxted, Matsuo, Medvedev, Miguel, Minier, Morello, Mura, Narita, Nascimbeni, Nguyen~Tong, Noce, Oliva, Palle, Palmer, Pancrazzi, Papageorgiou, Parmentier, Perger, Petralia, Pezzuto, Pierrehumbert, Pillitteri, Piotto, Pisano, Prisinzano, Radioti, R{\'e}ess, Rezac, Rocchetto, Rosich, Sanna, Santerne, Savini, Scandariato, Sicardy, Sierra, Sindoni, Skup, Snellen, Sobiecki, Soret, Sozzetti, Stiepen,
  Strugarek, Taylor, Taylor, Terenzi, Tessenyi, Tsiaras, Tucker, Valencia, Vasisht, Vazan, Vilardell, Vinatier, Viti, Waters, Wawer, Wawrzaszek, Whitworth, Yung, Yurchenko, Osorio, Zellem, Zingales, \& Zwart}]{ariel}
Tinetti, G., Drossart, P., Eccleston, P., {et~al.} 2018, Experimental Astronomy, 46, 135, \dodoi{10.1007/s10686-018-9598-x}

\bibitem[{{Tuomi} {et~al.}(2013){Tuomi}, {Jones}, {Jenkins}, {Tinney}, {Butler}, {Vogt}, {Barnes}, {Wittenmyer}, {O'Toole}, {Horner}, {Bailey}, {Carter}, {Wright}, {Salter}, \& {Pinfield}}]{tuomi13}
{Tuomi}, M., {Jones}, H.~R.~A., {Jenkins}, J.~S., {et~al.} 2013, \aap, 551, A79, \dodoi{10.1051/0004-6361/201220509}

\bibitem[{{van de Kamp}(1963)}]{kamp1963}
{van de Kamp}, P. 1963, \aj, 68, 515, \dodoi{10.1086/109001}

\bibitem[{{van de Kamp}(1969)}]{kamp1969}
---. 1969, \aj, 74, 757, \dodoi{10.1086/110852}

\bibitem[{{van de Kamp}(1975)}]{kamp1975}
---. 1975, \aj, 80, 658, \dodoi{10.1086/111791}

\bibitem[{{van de Kamp}(1982)}]{kamp1982}
---. 1982, Vistas in Astronomy, 26, 141, \dodoi{10.1016/0083-6656(82)90004-6}

\bibitem[{{Vanderburg} {et~al.}(2016){Vanderburg}, {Plavchan}, {Johnson}, {Ciardi}, {Swift}, \& {Kane}}]{Vanderburg2016}
{Vanderburg}, A., {Plavchan}, P., {Johnson}, J.~A., {et~al.} 2016, \mnras, 459, 3565, \dodoi{10.1093/mnras/stw863}

\bibitem[{{VanderPlas}(2018)}]{vanderplas2018}
{VanderPlas}, J.~T. 2018, \apjs, 236, 16, \dodoi{10.3847/1538-4365/aab766}

\bibitem[{Vehtari {et~al.}(2021)Vehtari, Gelman, Simpson, Carpenter, \& B{\"u}rkner}]{vehtari2021rank}
Vehtari, A., Gelman, A., Simpson, D., Carpenter, B., \& B{\"u}rkner, P.-C. 2021, Bayesian analysis, 16, 667

\bibitem[{{Vio} {et~al.}(2013){Vio}, {Diaz-Trigo}, \& {Andreani}}]{Vio2013}
{Vio}, R., {Diaz-Trigo}, M., \& {Andreani}, P. 2013, Astronomy and Computing, 1, 5, \dodoi{10.1016/j.ascom.2012.12.001}

\bibitem[{Walker(1971)}]{walker1971estimation}
Walker, A. 1971, Biometrika, 58, 21

\bibitem[{Welch(1967)}]{welch67}
Welch, P. 1967, IEEE Transactions on Audio and Electroacoustics, AU-15, 70

\bibitem[{{Wittenmyer} {et~al.}(2019){Wittenmyer}, {Clark}, {Zhao}, {Horner}, {Wang}, \& {Johns}}]{Wittenmyer2019}
{Wittenmyer}, R.~A., {Clark}, J.~T., {Zhao}, J., {et~al.} 2019, \mnras, 484, 5859, \dodoi{10.1093/mnras/stz290}

\bibitem[{{Wright}(2004)}]{wright04}
{Wright}, J.~T. 2004, \aj, 128, 1273, \dodoi{10.1086/423221}

\bibitem[{{Yu} {et~al.}(2017){Yu}, {Donati}, {H{\'e}brard}, {Moutou}, {Malo}, {Grankin}, {Hussain}, {Collier Cameron}, {Vidotto}, {Baruteau}, {Alencar}, {Bouvier}, {Petit}, {Takami}, {Herczeg}, {Gregory}, {Jardine}, {Morin}, {M{\'e}nard}, \& {Matysse Collaboration}}]{yu17}
{Yu}, L., {Donati}, J.~F., {H{\'e}brard}, E.~M., {et~al.} 2017, \mnras, 467, 1342, \dodoi{10.1093/mnras/stx009}

\bibitem[{{Zechmeister} \& {K{\"u}rster}(2009)}]{Zechmeister2009}
{Zechmeister}, M., \& {K{\"u}rster}, M. 2009, \aap, 496, 577, \dodoi{10.1051/0004-6361:200811296}

\end{thebibliography}
\bibliographystyle{aasjournal}
\appendix
    \section{Overview of the generalized Lomb-Scargle periodogram}\label{sec:gsl}

The generalized Lomb-Scargle periodogram \citep{Zechmeister2009}, which builds upon the work of \cite{lomb1976} and \cite{scargle1982}, is one of the most commonly used frequency-domain techniques in astronomy. 
For a time series of length $N$ with observations $y_i$, random errors $\sigma_i$ and observations times $t_i$, the GSLP is
\begin{equation}\label{eq:gsl}
\begin{split}
    P_{\mathrm{GLS}}(f) =\frac{1}{\sum_i^N w_i (y_i - \bar{y})^2 }\\
    \left\{\frac{\left[ \sum_{i}^{N} w_i(y_i - \bar{y})\cos{f(t_i - \tau)}\right]^2}{\sum_i^N w_i\cos^2{f(t_i-\tau)}-\left[\sum_i^N w_i\cos{f(t_i-\tau)}\right]^2} \right.+\\
    \left.\frac{\left[ \sum_{i}^{N} w_i(y_i - \bar{y})\sin{f(t_i - \tau)}\right]^2}{\sum_i^N w_i\sin^2{f(t_i-\tau)}-\left[\sum_i^N w_i\sin{f(t_i-\tau)}\right]^2}\right\},
\end{split}
\end{equation}
where $\bar{y}$ is the mean value of the observations $y_i$, $\tau$ is the time shift constant that makes evaluation of the periodogram independent of any offset in the observation times \citep{Press1987}, and $w_i$ are the normalized weights defined as
\begin{equation}
    w_i = \frac{1}{W\sigma_i^2}.
\end{equation}
$W$ is the sum of the inverses of the errors, i.e. $W=\sum_i  1/\sigma_i^2$ \citep{Zechmeister2009}. The parameter $\tau$ is defined in the GLSP as
\begin{equation}\label{eq:tau}
    \tan{(2f\tau)} = \frac{\sum w_i\sin{(2ft_i)} - 2\sum w_i \cos{f t_i} \sum w_i \sin{f t_i}}{\sum w_i \cos{(2ft_i)} - \left[\left(\sum \cos{f t_i}\right)^2 - \left(\sum \sin{f t_i}\right)^2\right]}.
\end{equation}

Assume now that the generating function of the time series is $y(t)$, which when sampled at times $t_i$ yields $y(t_i) = y_i$. If we knew $y(t)$ exactly, we would find its Fourier spectrum by projecting it onto the sum of weighted sines and cosines $g(t)$. Instead, we first approximate $g(t)$ by creating a regular mesh of length $L$ consisting of evenly spaced time points $\hat{t}_k$, where $k = 0,1,2,...,L$ and $L\geq N$. Lagrange interpolation gives
\begin{equation}
    \hat{g}(t) \approx \sum_k^L \gamma_k(t) g_k(\hat{t}_k).
\end{equation}
where $\gamma_k(t)$ are the interpolation weights. Now we interpolate the time series $y_i$ onto the regular mesh and multiply by $\hat{g}(t)$:
\begin{equation}\label{eq:press}
\begin{split}
    \sum_i^N y_i g(t_i ) \approx \sum_i^N y_i\left[\sum_k^L \gamma_k(t_i) g_k(\hat{t}_k)\right] = \\
    \sum_k^L \left[\sum_i^N y_i \gamma_k(t_i)\right]g_k(\hat{t}_k) = \sum_k^L \hat{y}_k g_k(\hat{t}_k),
\end{split}
\end{equation}
where $\hat{y}_k = \sum_i^N y_i \gamma_k(t_i)$ is the interpolated time series.

To find the GLSP, we define a new time series $h_i$, which is the product of the original mean-subtracted data and the normalized weights
\begin{equation}
h_i = w_i \left(y_i - \bar{y}\right).
\label{eq:weighted_time_series}
\end{equation}
We rewrite the numerators of the terms in Equation \eqref{eq:gsl} as
\begin{align}
    \sum_{i}^{N} w_i(y_i - \bar{y})\cos{f(t_i - \tau)} = \sum_i^N h_i \cos{f\left(t_i - \tau\right)}\label{eq:hi_1}\\
    \sum_{i}^{N} w_i(y_i - \bar{y})\sin{f(t_i - \tau)}=\sum_i^N h_i \sin{f\left(t_i - \tau\right)}.
    \label{eq:hi_2}
\end{align}
In this form, the cosine and sine terms play the role of the function $g(t)$ as in Equation \eqref{eq:press}. 
By performing the periodogram computation on a regularly spaced grid time grid, the \citep{Press1987} algorithm allows the observer to zero-pad the time domain.

\section{Additional experiments with varying time baselines}\label{app:timebaseline}

In Section \ref{sub:var_len} we showed that sinusoids with frequencies $f_1$ and $f_2$,
where $|f_1 - f_2| < 1.6\mathfrak{R}$, were either indistinguishable or had peaks at incorrect frequencies in the generalized Lomb-Scargle periodogram. In this section we show an additional two experiments that are readily applicable to exoplanet detection.

\begin{figure*}
     \centerline{\includegraphics[width=0.85\linewidth]{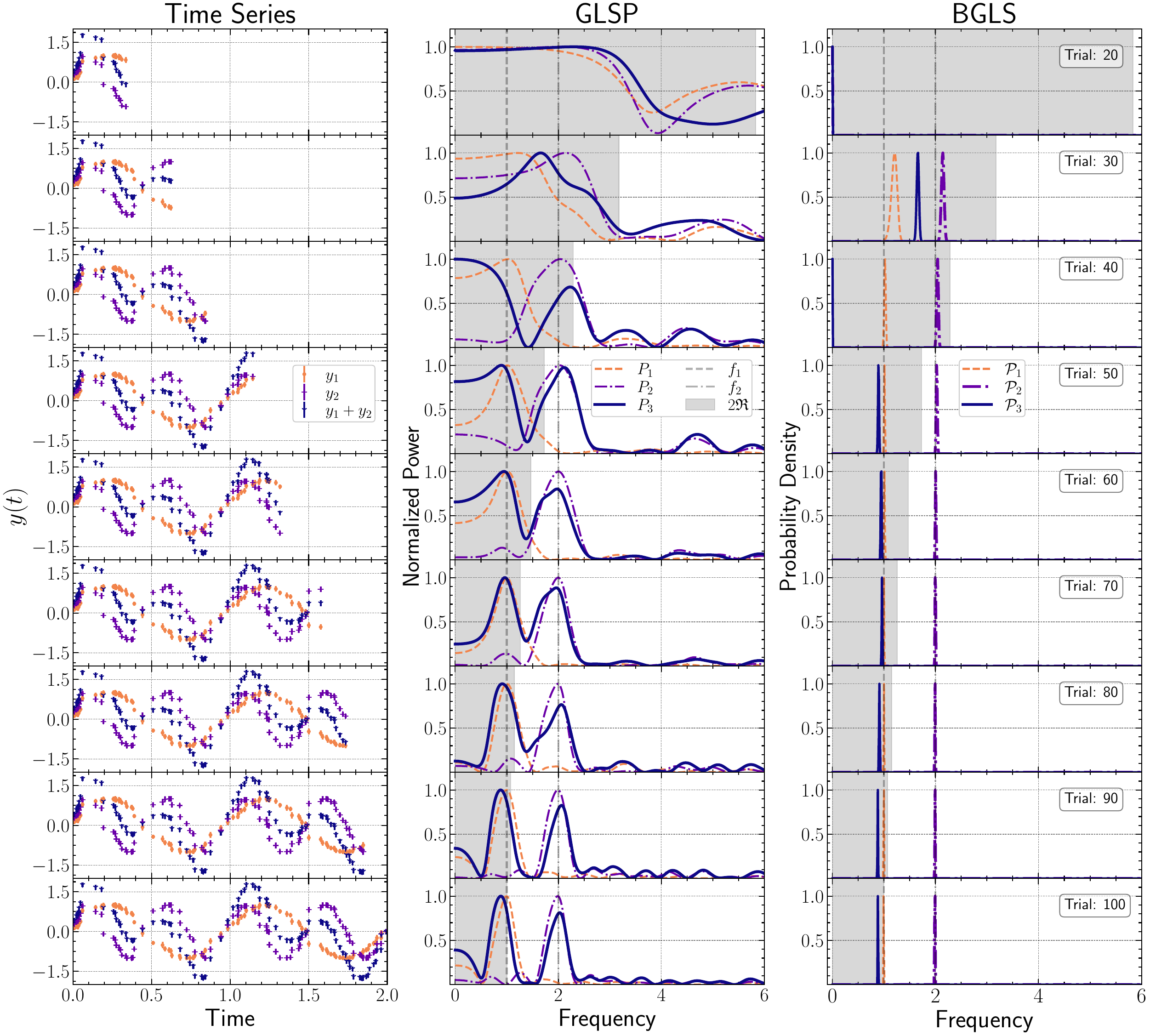}}
     \caption{{\bf Left column:} Unevenly spaced time series for $y_1$, $y_2$ and $y_3$, with the time baseline $T$ increasing from top to bottom. {\bf Middle column:} GLSP of each time series in the left column, illustrating the changes of the periodogram as the signal coverage increases. {\bf Right column:} BGLS periodogram of the time series in the left column, following the same rules as the middle column.}
    \label{fig:snapshot_uneventime}
\end{figure*}

The first experiment explores uneven time sampling by randomly choosing the timestamps from a uniform distribution on the interval $(0.01, 2)$. The functions $y_1$, $y_2$ and $y_3$ remain unchanged (Equations \ref{eq:y1}--\ref{eq:y3}). We retain identical first and last time stamps to the original experiment in Sect. \ref{sub:var_len} in order to preserve the same Rayleigh resolution. The results are shown in Fig.\ \ref{fig:snapshot_uneventime}. The uneven time sampling creates several spurious low-amplitude peaks in the GLSPs (middle column), but in general results remain consistent with the original experiment illustrated in Fig.\ \ref{fig:snapshot}. We observe that the Rayleigh criterion has be to satisfied for peaks to appear at the correct frequencies in $P_3$, the GLSP of $y_3$. The BGLS still has trouble identifying the correct frequencies for trials with $T < 1.2$ time units ($N < 60$).

\begin{figure*}
     \centerline{\includegraphics[width=0.85\linewidth]{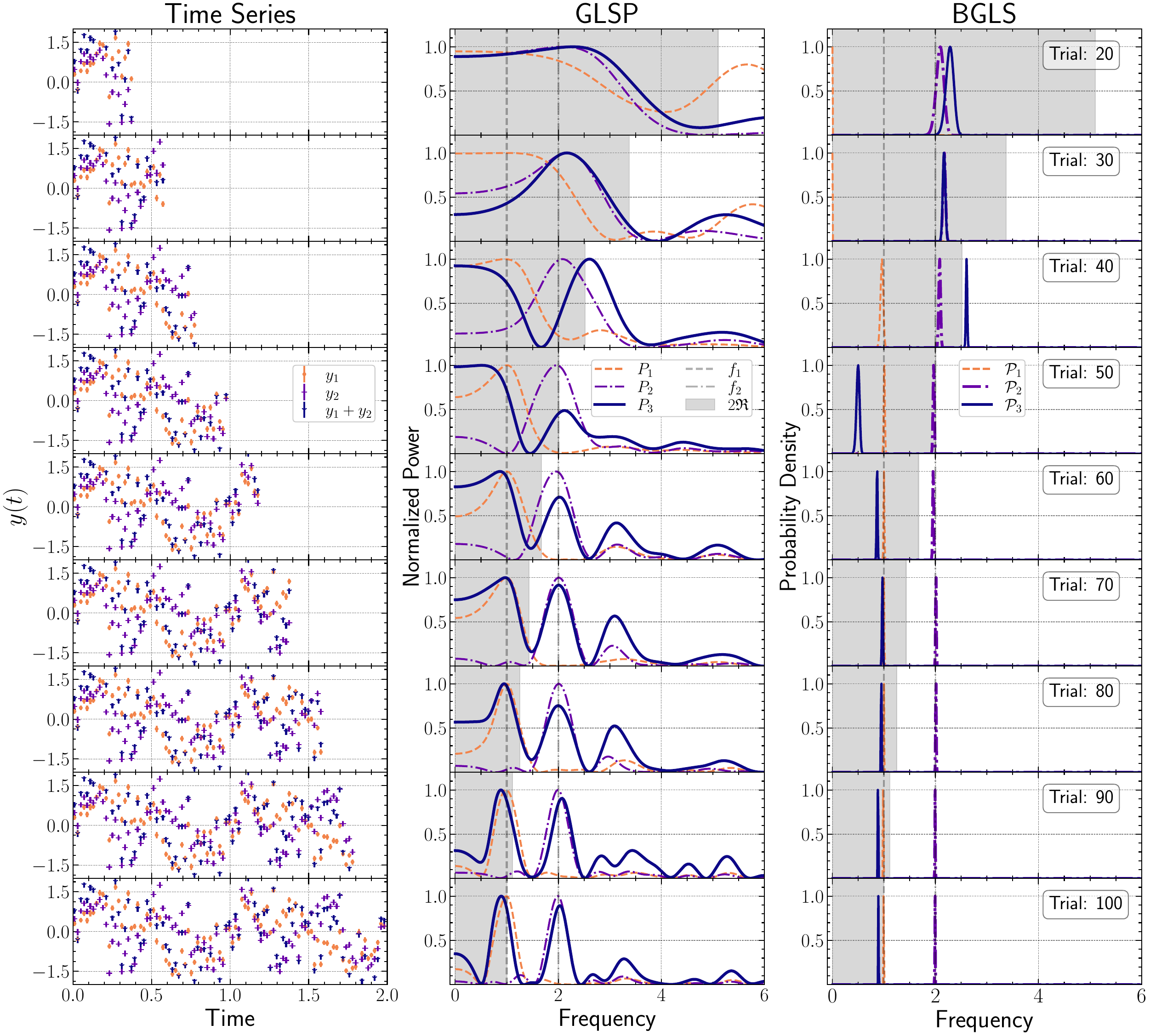}}
     \caption{{\bf Left column:} Time series $y_1$, $y_2$ and $y_3$ with the time baseline $T$ increasing from top to bottom. Simulated error bars are drawn from the distribution $\mathcal{N}(0, 0.5)$. {\bf Middle column:} GLSP of the time series in the left column, illustrating the changes of the periodogram as the signal coverage increases. {\bf Right column:} BGLS periodogram of the time series on the left column, following the same rules as the middle column.}
    \label{fig:snapshot_white_noise}
\end{figure*}

The second experiment involves varying the amplitude of the white noise added to the observations. The time series pictured in Figure \ref{fig:snapshot_white_noise} have error bars with amplitudes chosen from a narrow Gaussian distribution, $\mathcal{N}(0.1, 0.01)$. As a result, the error bars are nearly uniform. Here we randomly choose the error bars in Equations \ref{eq:y1} and \ref{eq:y2} from $\mathcal{N}(0, 0.5)$ in order to mimic large variations in the quality of the observations. The results are shown in Fig.\ \ref{fig:snapshot_white_noise}.
The added noise creates spurious peaks of appreciable amplitudes in the GLSPs (middle column). 
Furthermore, at trials 20--30 the higher-frequency peak in $P_3$ is not centered at the correct value, $f_2$. 
The BGLS have a few subtle differences from the original experiment in Sect. \ref{sub:var_len}. For example, the peak in $\mathcal{P}_2$ is not exactly centered at $f_2$ for trials 20--50.

\section{Signals reported for 55~Cnc, HD 99492 and Barnard's star}

We present in Table \ref{table:orbitalparameters} a summary of the signals reported in the literature for the RV cases analyzed in Sections \ref{subsec:55Cnc}, \ref{sub:hd99492} and \ref{sub:barnardstar}. The table shows the reported periods along with their reference, and we label the status of each detection ranging from confirmed, challenged or candidate signals.
{\renewcommand{\arraystretch}{1.5} 

\begin{table*}[ht]\label{table:orbitalparameters}
\centering
\caption{Signals reported in the literature for the stars 55~Cnc, HD 99492 and Barnard's star with their corresponding periods, status of the detection and reference for the period's value.}
\begin{tabular}{|c|c|c|c|c|}
\hline
\textbf{Star} & \textbf{Signal} & \textbf{Period [days]} & \textbf{Status} & \textbf{Reference} \\
\hline
\multirow{4}{*}{55~Cnc} & 55~Cnc b & $14.6516\pm 0.0001$ & Confirmed & \cite{bourrier2018} \\
& 55~Cnc c & $44.3989_{-0.0043}^{+0.0042}$ & Confirmed & \cite{bourrier2018} \\
& 55~Cnc d & $5574.2_{-88.6}^{+93.8}$ & Challenged & \cite{bourrier2018} \\
& 55~Cnc e & $0.73654737_{-1.44}^{+1.30}$ & Confirmed & \cite{bourrier2018} \\
& 55~Cnc f & $259.88\pm 0.29$ & Confirmed & \cite{bourrier2018} \\
& Activity Cycle & $3822.4_{-77.4}^{+76.4}$ & Challenged & \cite{bourrier2018} \\
\hline
\multirow{3}{*}{HD 99492} & HD 99492 b & $17.054\pm 0.003$ & Confirmed & \cite{kane2016} \\
& HD 99492 c & $4970 \pm744$ & Challenged & \cite{meschiari2011} \\
& HD 99492 c & $95.233_{-0.096}^{+0.098}$ & Confirmed & \cite{stalport2023} \\
\hline
\multirow{7}{*}{Barnard's star} & Barnard's star b & $232.8\pm0.4$ & Challenged & \cite{ribas2018} \\
& Barnard's star b & $3.1533 \pm 0.0006$ & Confirmed & \cite{gonzaleshernandez2024} \\
& Barnard's star c & $4.12$ & Candidate & \cite{gonzaleshernandez2024} \\
& Barnard's star d & $2.34$ & Candidate & \cite{gonzaleshernandez2024} \\
& Barnard's star e & $6.74$ & Candidate & \cite{gonzaleshernandez2024} \\
& Activity Cycle & $\sim 6600$ & Challenged & \cite{ribas2018} \\
& Activity Cycle & $3210^{+530}_{-430}$ & Confirmed & \cite{gonzaleshernandez2024} \\
\hline 
\end{tabular}
\end{table*}
}

\section{Posterior distributions of GP models of \textit{Kepler} Light Curves}\label{app:cornerplots}
\begin{figure*}
     \centerline{\includegraphics[width=0.85\linewidth]{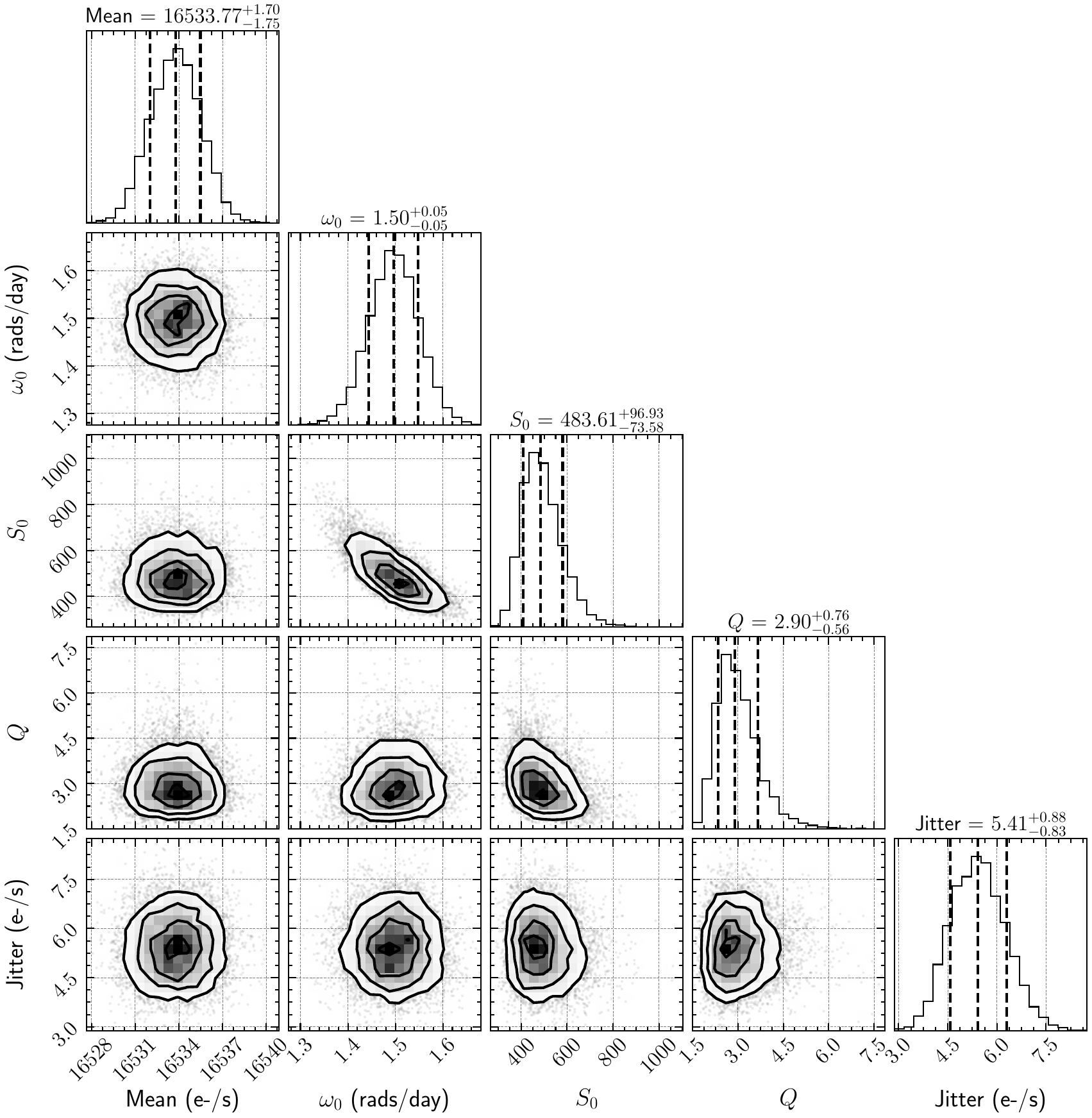}}
     \caption{Corner plot showing the posterior distribution for the GP model parameters for the light curve of KIC 891916. The vertical lines in the one dimensional histograms represent the 0.16, 0.5 and 0.84 percentiles.}
    \label{fig:cornerKIC891916}
\end{figure*}

\begin{figure*}
     \centerline{\includegraphics[width=0.85\linewidth]{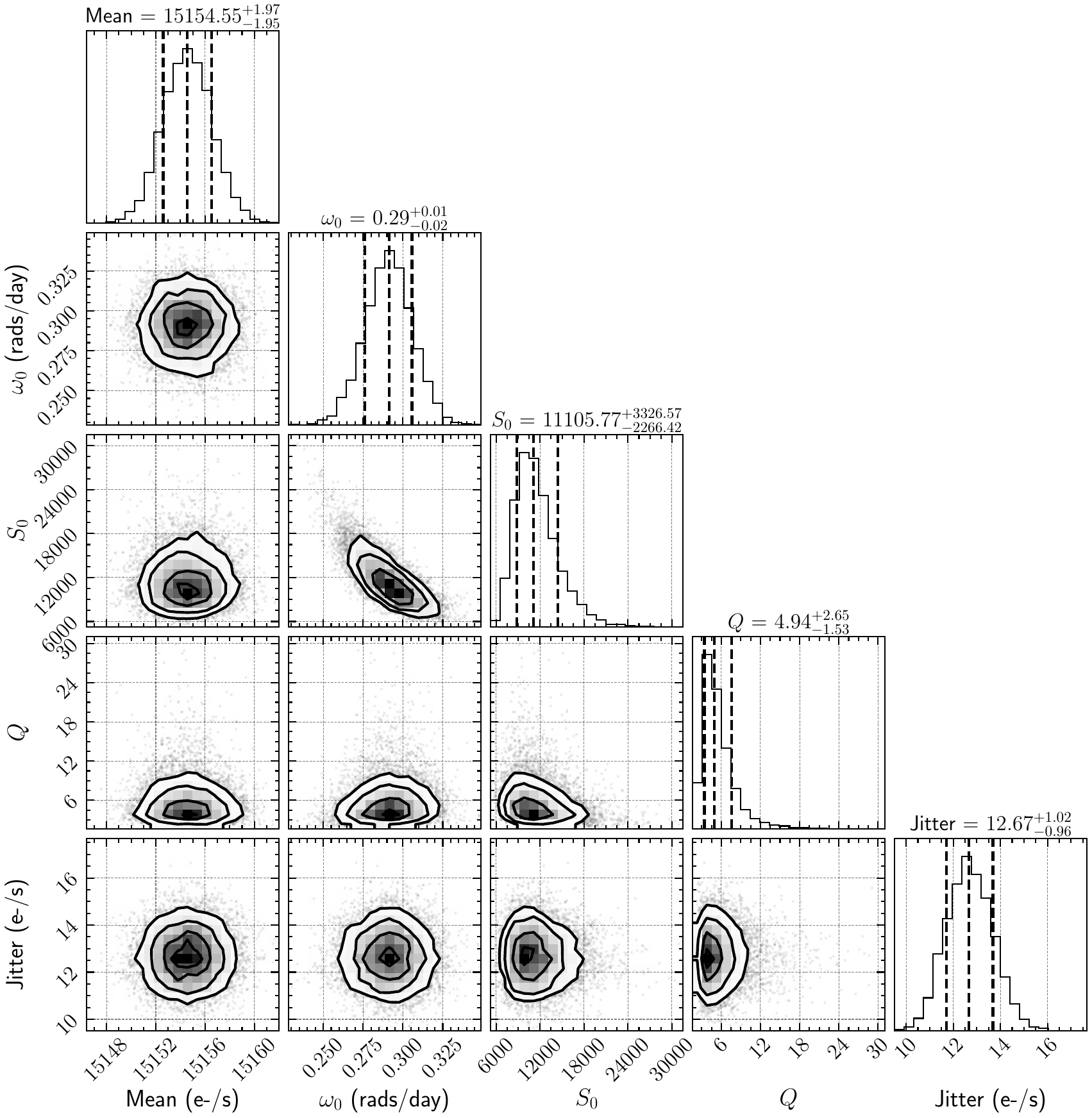}}
     \caption{Same as Figure \ref{fig:cornerKIC891916} for the case of KIC 1869783 light curve.}
    \label{fig:cornerKIC1869783}
\end{figure*}

This section presents the posterior distribution for the free parameters used in the GP model of the \textit{Kepler} light curves from Sec. \ref{subsec:diffrot}. The MCMC sampling was performed using 4 chains with 2000 draws per chain, making a total of 8000 draws from the posterior distribution. We used the potential scale reduction factor $\hat{R}$ diagnostic \citep{gelman1992inference} to test MCMC convergence in our posterior distributions. The value of $\hat{R}$ must be close to 1 to guarantee that the chains are well mixed and converged; if not, $\hat{R} > 1$. We used the package $\texttt{ArviZ}$ \citep{arviz} to compute $\hat{R}$ diagnostic from our posterior distributions. \texttt{ArviZ} computes \cite{vehtari2021rank}'s improved $\hat{R}$, which accounts for when the variance changes across the chains and when the chain has a heavy tail of values that are far from the mean. The value of $\hat{R}$ for the posterior distributions of all our parameters was equal to $1.0$, ensuring that the MCMC converged to a valid solution. The corner plots for our parameters' posterior distributions are shown in Figures \ref{fig:cornerKIC891916} and \ref{fig:cornerKIC1869783}.


\end{document}